\documentclass[a4paper,fleqn,usenatbib]{mnras}
\usepackage{amsmath}
\usepackage{ae,aecompl}
\usepackage{enumitem}
\usepackage{graphicx}
\usepackage{amssymb}
\usepackage{color}
\usepackage{threeparttable}
\usepackage{changes}
\usepackage{aas_macros}

\usepackage{url}
\usepackage{fancyref}
\usepackage{mVersion}

\title[Modelling nova populations in galaxies]{Modelling nova populations in galaxies}

\author[H.-L. Chen et al.]{Hai-Liang Chen$^{1,2,3,4}$\thanks{E-mail:
chenhl@ynao.ac.cn}, T. E. Woods$^{1,5}$, L. R. Yungelson$^{6}$, M.
Gilfanov$^{1,7,8}$, Zhanwen Han$^{2,3}$\\
  $^{1}$Max Planck Institute for Astrophysics, Karl-Schwarzschild-Str. 1,
Garching b. M{\" u}nchen 85741, Germany\\
  $^{2}$Yunnan Observatories, Chinese Academy of Sciences, Kunming, 650011,
China\\
  $^{3}$Key Laboratory for the Structure and Evolution of Celestial Objects, 
Chinese Academy of Sciences, Kunming 650011, China\\
  $^{4}$University of Chinese Academy of Sciences, Beijing 100049, China\\
  $^{5}$Monash Centre for Astrophysics, 19 Rainforest Walk, Monash University, VIC, 3800, Australia\\
  $^{6}$Institute of astronomy, RAS, 48 Pyatnitskaya Str., 119017 Moscow, Russia\\
  $^{7}$Space Research Institute of Russian Academy of Sciences, Profsoyuznaya
84/32,117997 Moscow, Russia\\
  $^{8}$Kazan Federal University, Kremlevskaya str.18, 420008 Kazan, Russia\\
}

\date{Accepted XXX. Received YYY; in original form ZZZ}

\begin{document}
\label{firstpage}
\pagerange{\pageref{firstpage}--\pageref{lastpage}}
\maketitle

\begin{abstract}

Theoretical modelling of the evolution of classical and recurrent novae plays an important role in studies of binary
evolution, nucleosynthesis and accretion physics. However, from a theoretical perspective the observed statistical 
properties of novae remain poorly understood. In this paper, we have produced model populations of novae using a hybrid 
binary population synthesis approach for differing star formation histories (SFHs): a starburst case (elliptical-like galaxies), a constant star formation rate case 
(spiral-like galaxies) and a composite case (in line with the inferred SFH for M31). We found that the nova rate at 10\;Gyr in an
elliptical-like galaxy is $\sim 10-20$ times smaller than a
spiral-like galaxy with the same mass. The
majority of novae in elliptical-like galaxies at the present epoch are characterized by low mass white dwarfs (WDs), long decay times, relatively faint absolute magnitudes and long recurrence periods. In contrast, the majority of novae in
spiral-like galaxies at 10\;Gyr have massive WDs, short decay
times, are relatively bright and have short recurrence periods. The
mass loss time distribution for novae in our M31-like galaxy is
in agreement with observational data for Andromeda. However, it is possible
that we underestimate the number of bright novae in our model. This may
arise in part due to the present uncertainties in the appropriate bolometric correction for novae.

\end{abstract}


\section{Introduction}
\label{sec:int}

As a subclass of accreting white dwarfs, novae are important objects
for the study of binary evolution and nucleosynthesis \citep[see][for
  a review]{patt84,gtws98}.
Given that WDs undergoing nova eruptions may gain mass
during accretion, it has been suggested that they may be the progenitors
of type Ia supernovae in some variants of the single degenerate scenario
\citep[e.g.][]{sss88,yltt+96,hk01}. Novae are believed 
    to be important sources of some nuclides, such as $^{7}$Li, $^{15}$N, 
$^{17}$O, $^{22}$Na and $^{26}$Al \citep[e.g.][]{sts78,hjci96,gtws98,kct00} which in some cases have been 
actually observed \citep{tsna+15,idmm+15,tsna+16}. Although novae have been
widely studied, many questions remain unclear. In recent
years, an increasing number of observational studies of novae have become available. 
It is vital to compare our present theoretical understanding with observations, in particular the properties of nova populations on the whole.

It is accepted that novae occur in an accreting WD binary with
accretion rates below the stable burning regime, whereby the H-rich material
burns unstably on the surface of the WD
\citep[e.g.][]{mest52a,mest52b,kraf64,gw67,ty72,pz78,sat79, te79}. But note, {\bf \citet{iss13}} found that
in the high accretion rate regime which is conventionally considered as corresponding to 'stable' H-burning,
the latter actually proceeds in small-scale short-timescale flashes.
In order to understand the process underlying nova explosions, considerable effort
has been made in producing simulations of nova explosions 
\citep[e.g.][]{stsk72,pss78,pss79,sat79,pk95,ypsk05}. During
the accretion phase, first, H-rich material will be accumulated on the
surface of the WD and gradually compressed by more accreted
material. The H-rich shell will be heated by the compression and
undergo a thermonuclear runaway (TNR), when the pressure at the bottom of
the accreted envelope becomes sufficiently high and the degeneracy is
lifted.  The TNR will lead to ejection of the
accreted mass. If the accretion rate is close to the stable burning
regime, the nova explosion is relatively weak and only a fraction of the
accreted mass will be ejected. If the accretion rate is low enough,
nova explosions are strong and the WD will be eroded.  \citet{pk95} and
\citet{ypsk05} have simulated a large number of multicycle nova
evolution models. They found that the properties of novae
(e.g. ignition mass, maximum luminosity) are mainly determined by the
WD mass, accretion rate and the interior temperature. They show
different characteristics of novae for accreting WDs with different
parameters within the possible ranges.  Based on the results of these
simulations, many general properties of individual nova can be
understood.

In the past three decades, observations of novae in different Hubble
types of galaxies have been undertaken
\citep[e.g.][]{ctjf+90,drbl94,scp00,fcj03,ws04,csm08,fshm12}. These
observations are very helpful to study nova properties in different
stellar populations. Previous studies \citep[e.g.][]{duer90,dblo92}
suggested that there are two kinds of nova population: the disk novae
and bulge novae. The disk novae are fast and may harbour more massive WDs,
compared with the bulge novae \citep{dl98}. 
Some observational studies \citep[e.g.][]{ctjf+90,scp00,fcj03,ws04} have suggested that
luminosity-specific nova rates may not evolve strongly with the Hubble
type of galaxies and the average luminosity-specific nova rate is around
$2\times10^{-10}\,{L_{\rm K,\odot}}^{-1}\,{\rm yr}^{-1}$. However, several
studies indicate that some galaxies, such as the Magellanic Clouds, M33 and others, may have higher luminosity-specific nova rates \citep[e.g.][]{drbl94,dell02,ns05,as14,sdlz+16}. 

Regarding individual galaxies, novae in M31 have been intensively studied
in the past \citep[e.g][]{arp56,rosi64,rosi73,cfnj+87,rcdd89,si01,dbkn+06}.
Some earlier results suggested a nova rate in M31 of 
$\sim 20-40\;{\rm yr}^{-1}$ (see \cite{si01} and their table 1). \citet{dbkn+06} performed
a careful analysis of the distribution and the incompleteness of novae
in their survey, and derived a nova rate of $65^{+16}_{-15}\,{\rm
  yr}^{-1}$. These studies have also provided other key parameters in describing nova
properties (e.g. decay time, peak magnitude), which can be directly
compared with theoretical results.

With the binary population synthesis method, \citet{ylt97} found that the
luminosity-specific nova rate should be significantly higher in
younger stellar populations compared with older stellar
populations. The typical WD mass of novae in young stellar populations
is larger than in old stellar populations. In addition, \citet{nmd04}
modelled the Galactic nova population (assuming a constant SFR and 100\% initial 
binarity) and found that their derived nova rate and
orbital period distribution are in agreement with observations.

In this paper, with a hybrid binary population synthesis method, we
model the nova population for galaxies with two representative star
formation histories (SFHs), i.e. starburst and constant star formation
rate (SFR). Moreover, in order to compare with nova surveys of M31, we have computed a composite model with a star formation history consistent with that which is presently inferred for M31 \citep[see][and discussion below]{rysh04}. We will present our predicted nova
rates, as well as the characteristic distribution of novae for different models,
such as the WD mass, peak magnitude, mass loss time, and compare these
results with observations.

The paper is structured as follows. In section~\ref{sec:bps}, we
introduce our binary population synthesis approach and describe how we
calculate the values of various nova properties in our simulation models.
In section~\ref{sec:resu}, we mainly show the results of these
simulations and compare these results with observations. We have a further
discussion in section~\ref{sec:disc}, before summarizing our results in
section~\ref{sec:con}.

\section{Binary population synthesis}
\label{sec:bps}

In this paper, we adopt a hybrid binary population synthesis approach,
which was first introduced in \citet{cwyg+14}. In this
approach, first, we use the \textsc{bse} code \citep{htp02} to generate WD
binaries with non-degenerate donors (i.e. main sequence (MS),
Hertzsprung gap (HG), red giant (RG) stars). Then we follow the
evolution of these WD binaries with a grid of detailed evolutionary
tracks computed with \textsc{mesa} code
\citep{pbdh+11,pcab+13}. Compared with other binary population
synthesis approaches, this method allows a careful treatment of the
second mass transfer phase.  Here we briefly summarize the main
assumptions and ingredients in our calculation.

\subsection{BSE calculation}

We adopt the initial mass function (IMF) of \citet{krou01} for the
primary mass ranging from $0.1\,\rm M_{\odot}$ to $100\,\rm
M_{\odot}$. We take a flat mass ratio distribution \citep{kpty79} and
a flat distribution in logarithmic space for binary separations in the
range between $10$ and $10^{6} \mathrm{R}_{\odot}$. With respect to
the binary fraction, observations indicate that it may depend on the
binary parameters \citep{kbgp+09,kh09,sddl+12}. Following the
suggestion of \citet{vnvt+13}, we adopt the following formula for the
binary fraction ($f_{\rm b}$):
\begin{equation}
\label{equ:bin_fra}
f_{\rm b} = 0.50+0.25{\rm log}_{10}(M_{1}),
\end{equation} 
where $M_{1}$ is the primary mass. This is different from our previous papers
\citep{cwyg+14,cwyg+15}, in which we adopt a constant binary fraction, i.e. 50\%.
However, this does not significantly influence our results.

With the \textsc{bse} population synthesis code \citep{htp02}, we computed
the evolution of around $2\times10^{6}$ binary systems, a fraction of
which may evolve into a binary system consisting of a WD and a non-degenerate
donor that later stably overflows Roche lobe. We follow the evolution of these systems until the WD binaries
become semi-detached and obtain the binary parameters (i.e. WD mass,
donor mass, orbital period) at the onset of mass transfer.

It is worth noting that we include different types of donors (i.e. MS,
HG, RG donors) in our calculation. In previous studies of nova
populations (e.g. \citealt{nmd04}), only MS donors are
included. However, \citet{dbhh+14} have shown that some recurrent
novae harbour HG and RG donors, though some of these may reside in 
symbiotic binaries. Therefore, for the first time, we include a
relatively complete set of progenitors of novae binaries in a population synthesis
study.

\subsection{Binary evolution calculation}

In order to follow the evolution of these WD binaries, we computed the
evolution of a grid of WD binaries with varying initial parameters
using the detailed stellar evolutionary code \textsc{mesa}
\citep{pbdh+11,pcab+13}. In this grid, the WD mass ranges from
$0.50\,\rm M_{\odot}$ to $1.35\,\rm M_{\odot}$ with a step $\Delta{M}
= 0.10\,\rm M_{\odot}$. We ignore He WDs in our calculation.\footnote{
  \citet{nmd04} computed the nova rates from He WDs by extrapolating
  the results of \citet{pk95} and found that He WDs contribute a
  very small fraction ($\lesssim 5\%$) to the total nova rates.}
The donor mass ranges from $0.10 \,\rm M_{\odot}$ to
$13.5\,\rm M_{\odot}$ with a step $\Delta{M} = 0.05 \,\rm
M_{\odot}$. The orbital period ranges (with step size $\Delta{\rm log}_{10}P_{\rm orb} = 0.1$)
from 1000 days down to the minimum period for which the binary separation yields 
mass transfer which begins on the zero-age main sequence. With the
binary parameters at the onset of mass transfer from our \textsc{bse}
calculations, we can select the closest track in the grid of
\textsc{mesa} calculation and follow the evolution of any WD binaries.

In the \textsc{mesa} calculation, we adopt the limits for the stable
burning regime from \citet{wbbp13}. We assume that the excess mass
will be lost in the form of an optically thick wind \citep{hkn96} and take
away the specific angular momentum of the WD, if the accretion rate is
larger than the maximum stably burning rate. If the accretion rate is
in the stably burning regime, we assume that no mass will be lost. If
the accretion rate is below the stable burning regime, we adopt the H
burning retention efficiency from \citet{ypsk05}, 
as approximated in \citet{cwyg+14} (see their Eq. 6). Regarding the He burning retention 
efficiency, this suffers from considerable uncertainties 
\citep[e.g.][]{hkn99,kh04,pty14,hpks15}. Here we simply use the 
prescription from \citet{hkn99}. The angular momentum lost due to  mass loss during nova explosion is rather uncertain \citep[see e.g.][]{lgr91,nsrt+16,szw16}.
Here we assume that the lost mass takes away the specific angular momentum of 
the WDs.

In our calculation, we adopt an initial hydrogen abundance $X = 0.70$,
helium abundance $Y = 0.28$ and metallicity $Z = 0.02$.

\subsection{Calculation of the nova rate}

In our calculations, we do not compute the detailed evolution of the
WDs.  We consider the WD as a point mass and compute its mass
evolution. From our binary evolution calculations, for any given WD
binary evolutionary track, we know the accretion rate and WD
mass. 
\citet{ypsk05} computed multicycle nova models for WDs under differing conditions, and showed the dependence of nova properties on WD mass,
accretion rate and WD temperature (see their tables 2 and 3). 
Under certain assumptions on the interior temperature of the WDs, we can then find 
the characteristics of Novae, such as the ignition and ejected mass, by
interpolating their tables. The results of \citet{tb04} indicate that, for WD masses and 
accretion rates typical of cataclysmic variables, equilibrium WD core 
temperatures are smaller than $1.0\times10^{7}$\,K. However, 
there is no complete computation similar to \citet{ypsk05} for different WDs with
low temperatures. Therefore, we use $T_{\rm c} = 1.0\times 10^{7}$\;K
in our calculations. In section~\ref{sec:disc}, we will discuss its influence
on our results.

\subsection{Common envelope evolution}

For the treatment of common envelope (CE) events, we adopt the
$\alpha$-formalism \citep{webb84,deko90} and use the fitting formula
from \citet{lvk11} for the binding energy parameter $\lambda$. The
efficiency $\alpha$ suffers from considerable uncertainty
\citep[see][for a review]{ijcd+13}. \citet{dkk12} found that the value
of $\alpha$ should be $\le 1.0$ and it may decrease with increasing WD
mass and secondary mass. By reconstructing the evolution of post CE
binaries, \citet{zsgn10} constrain the value of $\alpha$ to be in the
range of $0.2-0.3$.  A similar result is found by \citet{rt12} using
hydrodynamic simulations.  \citet{dkw10} found that a global value of
$\alpha >0.10$ provides a good agreement with observations of post
CE binaries. Therefore, in our calculation, we adopted $\alpha =
0.25$. We will discuss the influence of this value on our results in
section~\ref{sec:disc}.

By default, we adopt criteria for dynamically stable mass loss from \citet{hw87} and \citet{webb88}
(hereafter the HW criteria) for binaries with giant stars. However, some
detailed binary evolution calculations \citep[e.g.][]{hpmm+02,ch08}
indicate that this criteria may predict too low a critical mass ratio,
and that the critical mass ratio should depend on the evolutionary phase of
the donor star, mass, and adapted mode of angular momentum loss from the
system. \citet{wi11} \citep[see also][]{php12} demonstrated that the
difference is due to consideration of the superadiabatic outer surface layer, whose
typical thermal timescale may be smaller than the timescale for mass transfer. The muted response of this surface layer of the donor in response to mass loss (relative to mass loss directly from the convective envelope, as in polytropic models), allows for stable mass transfer at higher mass ratios than previously considered. Recently, \citet{pi15} found that the critical
mass ratio may vary from 1.5 to 2.2.

\subsection{Binary population synthesis models}

With the uncertainties in the treatment of CE events in mind, we computed three
different models: a025 model, a025qc15 model, a025qc17 model. In all
of these models, we use $\alpha = 0.25$. In the first model, we use the HW
criteria for giant stars. In the other models, we use critical mass
ratios $q_{\rm c} = 1.5$ and $1.7$ for giant stars, respectively. However,
we should emphasize that the critical ratio should not be a fixed
value and varies as a function of other stellar parameters,
as mentioned above.

In each model, we consider three different SFHs, which may roughly
represent three kinds of galaxies. (I) Elliptical-like galaxies: all
stars (with initial total mass $10^{11}\,\rm M_{\odot}$) are formed at
$t = 0$. (II) Spiral-like galaxies: a constant SFR for 10 Gyr with total
stellar mass $10^{11}\,\rm M_{\odot}$ formed. (III) M31-like galaxies:
In order to compare wtih observations of novae in M31, we make
a composite model with a SFH outlined as follows. \citet{rysh04}
simulated the formation and SFH of disk galaxies, which \citet{obsd+06} have demonstrated to be consistent with the observationally derived SFH of M31.  Given
that the total stellar mass of M31 is around $1.1\times10^{11}\,\rm
M_{\odot}$ \citep{babe+07}, we adopt the SFH from \citet{rysh04} as
the SFH of M31 and rescale the total stellar mass to
$1.1\times10^{11}\;{\rm M_{\odot}}$. In this model, the SFR increases
from $t = 0$\,Gyr and peaks around $t = 2.0$\, Gyr, and then declines
by around one order of magnitude by 10 Gyr.

\begin{table*}
	\centering
	\caption{The current nova rates (i.e. at 10 Gyr) for different kinds of galaxies in different 
models. 
    The total stellar mass for elliptical-like and spiral-like galaxies is $10^{11}\;{\rm M}_{\odot}$ and it is
    $1.1\times10^{11}\;{\rm M}_{\odot}$ for M31-like galaxies. The present nova rate of M31 galaxy 
is 
    around $97\,{\rm yr}^{-1}$ (see text). }
	\label{tab:model_tab}
	\begin{tabular}{lcccccc}                                   
		\hline
		model  & $\alpha$ & $q_{\rm c}$ & WD temperature ($T_{\rm c}$)   & nova rate (${\rm yr}^{-1}$)  & 
nova rate (${\rm yr}^{-1}$) & nova rate (${\rm yr}^{-1}$) \\
               &          &             &     ( $10^{7}$\,K)             & (elliptical-like) & (
spiral-like) & (M31-like) \\
        (1)    &  (2)     & (3)         &        (4)                     & (5)         & (6)        
    &  (7) \\       
		\hline
		a025     & 0.25   & HW criteria &         1                      & 20          & 413             &
 158  \\
		a025qc15 & 0.25   & 1.50        &         1                      & 14          & 283             &
 105   \\
		a025qc17 & 0.25   & 1.70        &         1                      & 12          & 207             &
 80   \\
        a050     & 0.50   & HW criteria &         1                      & 13          & 468        
     & 160  \\
        a025     & 0.25   & HW criteria &         3                      & 42          & 462        
     & 198   \\
		\hline
	\end{tabular}
    \begin{tablenotes}
    \item {\bf Notes.} (1) model name. (2) $\alpha$ value.  (3) CE criteria for binaries with giant 
stars at the first mass transfer phase.  (4) WD interior temperature (5)-(7) nova rates at 10 Gyr.
    \end{tablenotes}
\end{table*}

\section{Results}
\label{sec:resu}

\subsection{Evolution of nova population with stellar ages}
\label{subsec:evo_nov}

\begin{figure*}    
	\includegraphics[width=\textwidth]{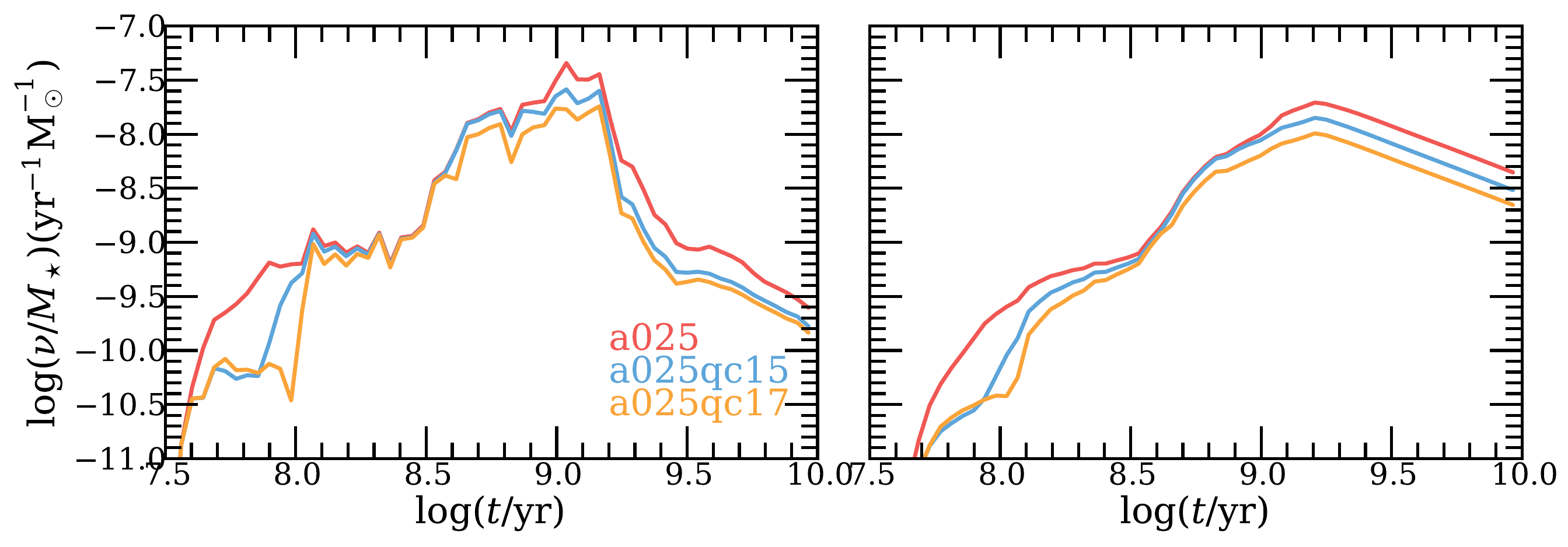}
    \caption{Evolution of mass-specific nova rates for elliptical-like galaxies
     (left panel) and spiral-like galaxies (right panel) in
      different models (see table~\ref{tab:model_tab}). The red, blue, orange colours are for a025,
      a025qc15, a025qc17 model, respectively.  }
    \label{fig:rnova}
\end{figure*}

\begin{figure*}  
	\includegraphics[width=\columnwidth]{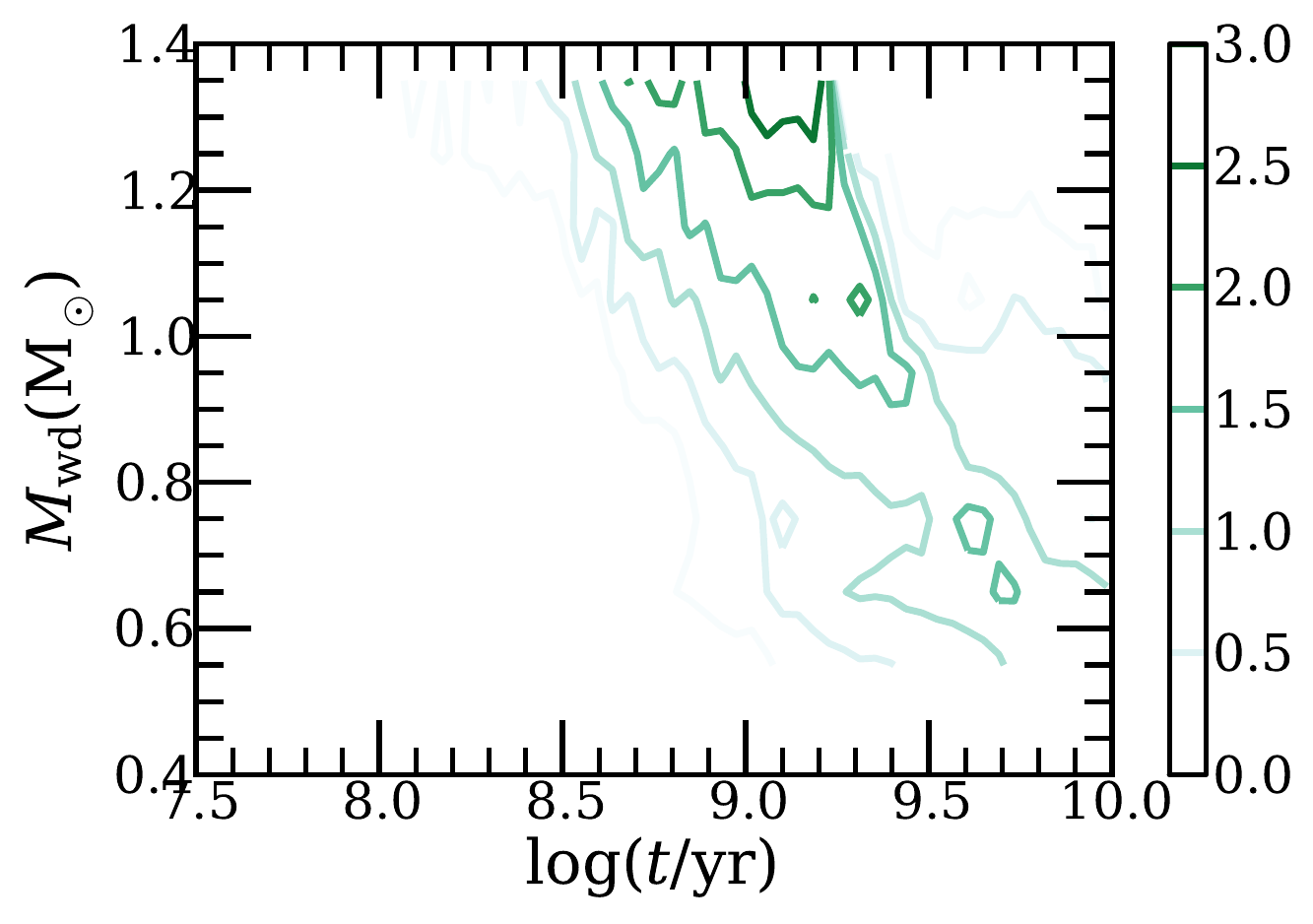}
	\includegraphics[width=\columnwidth]{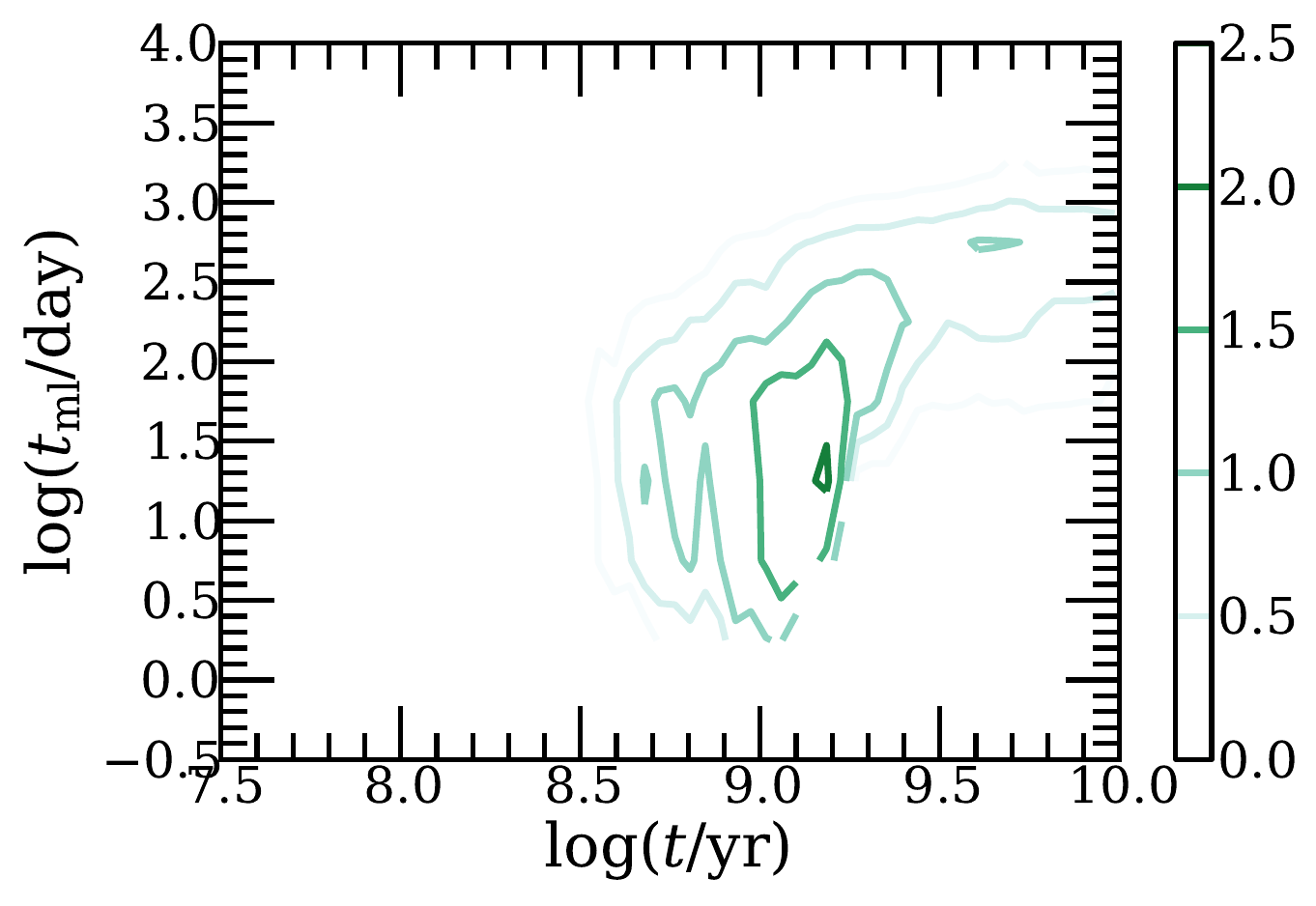}
	\includegraphics[width=\columnwidth]{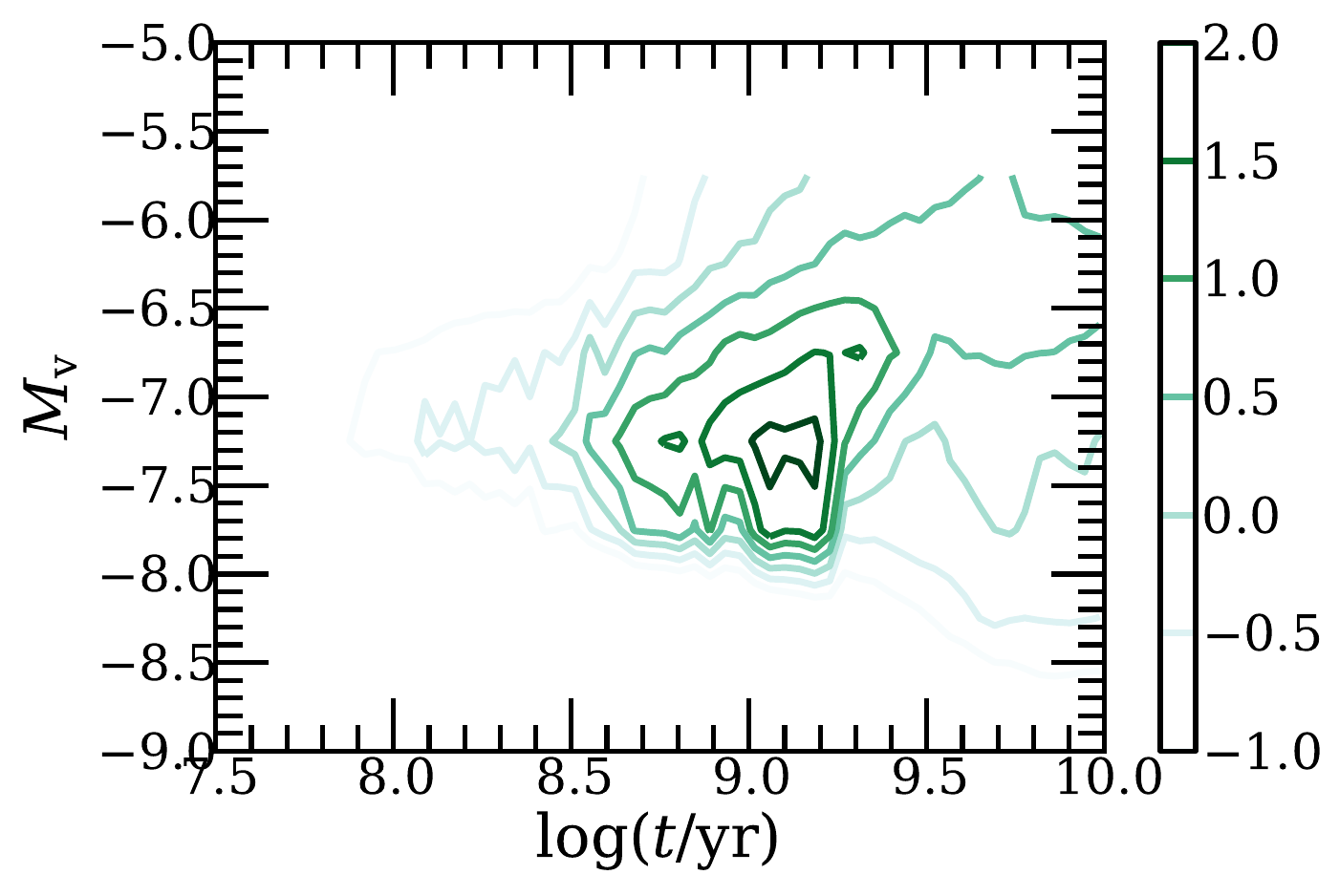}
	\includegraphics[width=\columnwidth]{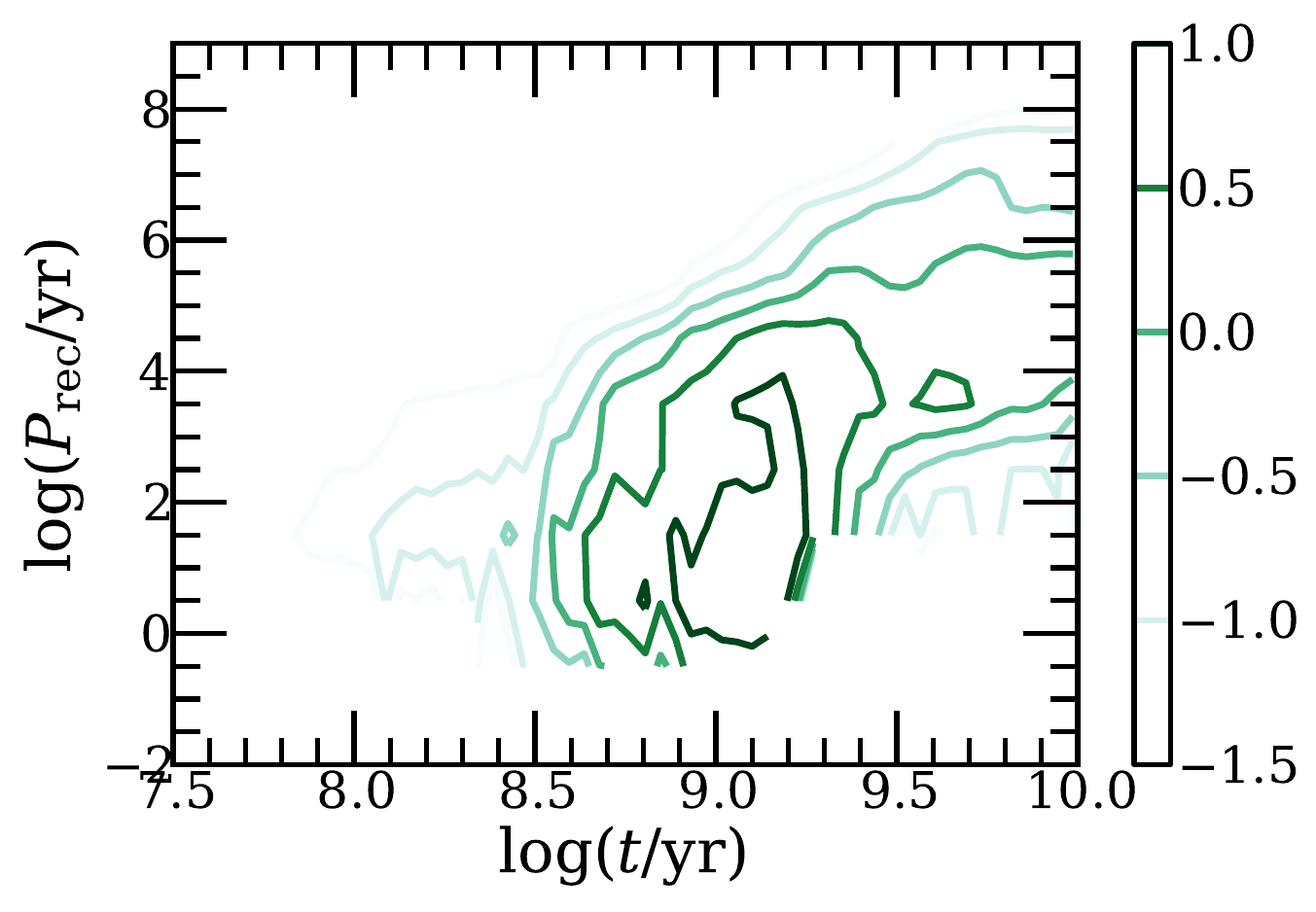}
    \caption{Isodensity contours for nova properties at different stellar ages for 
    elliptical-like galaxies in a025 model. The values of different contours are for 
    $(\partial^2 N/\partial {\rm log}t \partial M_{\rm WD})/M_{\star}$(upper left),  
    $(\partial^2 N/\partial {\rm log}t \partial {\rm log}t_{\rm tml})/M_{\star}$ (upper right), 
    $(\partial^2 N/\partial {\rm log}t \partial M_{\rm v})/M_{\star}$ (lower left),  
    $(\partial^2 N/\partial {\rm log}t \partial {\rm log}P_{\rm rec})/M_{\star}$ (lower right) in 
    logarithm. $N$ is the number of nova events and $M_{\star}$ is the stellar mass of the galaxy. 
    }
   
    \label{fig:contour_diff_times}
\end{figure*}

In Fig.~\ref{fig:rnova}, we show the evolution of nova rates for 
elliptical-like and spiral-like galaxies in different models. 
The mass-normalized nova rates reach maximum around $1$\;Gyr in elliptical-like 
galaxies and around $2$\;Gyr in spiral-like galaxies. 
After the maximum, it declines by $\sim$2 orders of magnitude in elliptical-like galaxies and by a
factor of $\sim$4 in spiral-like galaxies by the age of 10 Gyr. 
A similar behaviour in elliptical galaxies is also found in \citet{mrpd03}. In
elliptical-like galaxies, at $\sim$1\;Gyr, the typical donor mass
of WD binaries is around 2\;$\rm M_{\odot}$. For these binaries, WDs will
efficiently accumulate mass during a thermal timescale mass
transfer. These massive WDs have frequent outbursts. With increasing stellar 
age, the typical donor mass will decrease. The WD can not 
efficiently accumulate mass and the typical WD mass decreases. Therefore,
the nova rate will decrease. In spiral-like galaxies, novae form with 
a typical delay time around 1\;Gyr. Therefore, after $\sim$2\;Gyr, the 
nova rate is almost constant while the stellar mass continues to increase linearly
with time. Therefore, the mass-specific nova rates at old ages in the right 
panel of Fig.~\ref{fig:rnova} decrease as $t^{-1}$. 

Compared with our a025 model, the nova rates in a025qc15 and a025qc17 are
smaller. In a025qc15 and a025qc17 models, the critical mass ratios for a CE in
binaries with giant stars are larger than when using the HW criteria. Therefore,
it will become more difficult for binaries to enter a CE and fewer accreting WD
binaries will be produced. Consequently, the nova rates decrease as
this critical mass ratio increases. 

The nova rates at 10 Gyr in different models for different kinds of galaxies are shown in
table~\ref{tab:model_tab}. For an elliptical-like galaxy with
$10^{11} \,\rm M_{\odot}$, the nova rate is around $10-20\,{\rm
  yr}^{-1}$. For a spiral-like galaxy with the same mass, the nova rate
  is around 10-20 times larger. These results are in line with 
\citet{mrpd03} who found that the luminosity-specific nova rates in elliptical galaxies 
are lower than in spiral and irregular galaxies. \citet{ylt97} came to a similar conclusion using a binary population 
synthesis method. In M31-like galaxies, the
nova rate is around $80-160\,{\rm yr}^{-1}$. \citet{dbkn+06} detected
20 classical novae in M31 and deduced a global nova rate of
$65^{+16}_{-15}\,{\rm yr}^{-1}$ based on a thorough analysis of the
completeness of novae in their survey. However, as they discussed, the
novae with the shortest decay times $t_{2}$ or $t_{3}$ (the decline time of
the optical light curve from the peak by 2 (3) magnitudes) in their
sample are likely incomplete because of insufficient cadence of the survey.
In order to get a reliable nova rate for M31, \citet{sg14} and \citet{sgwb15} 
combined the data from observations of novae with $t_{2} < 20$ days by
\citet{arp56} with the data of novae with $t_{2} > 20$ days from
\citet{dbkn+06}, taking the completeness into consideration. 
In the incompleteness analysis of Arp's survey, they ran Monte-Carlo simulations 
and used a template nova light curve and an observed maximum magnitude-rate of decline
relation. Then they can get the fraction of detected novae as a function of $t_{2}$
(see Appendix of \citet{sgwb15} for more details). Regarding the 
incompleteness of Darnley's survey, we use a completeness value of $\approx 23\%$ for 
correction based on the result of \citet{dbkn+06} (see their table 3).
The nova rate is computed as follows.
$
\dot{N}(t_{2}) = \begin{cases}
    \frac{N}{\eta\times{1.342}\times6.02\times10^{10}}, & \text{for Arp's data}\\
    \frac{N}{\eta\times{2.830}\times{(0.47\times11\times10^{10})}}, & \text{for Darnley's data}.
                 \end{cases}
$\\
where $\dot{N}(t_{2})$ is in unit of ${\rm yr}^{-1}\;{\rm M_{\odot}}^{-1}$, $N$ is the observed numbers 
of sources at $t_{2}$ and $\eta$ is the completeness. The duration of the two surveys are 1.342 yr and 2.830 yr respectively. The masses enclosed by the field
of view in the two surveys are $6.02\times10^{10}\;{\rm M_{\odot}}$ and $5.17\times 10^{10}\;{\rm M_{\odot}}$,
respectively (see \citet{sgwb15} for more details). Then we obtain a global nova rate $97\,{\rm yr}^{-1}$ \footnote{In \citet{sgwb15}, they 
found a nova rate of $106\;{\rm yr}^{-1}$, corresponding to the $\theta = 1.0$ case 
of \citet{dbkn+06}. Here, we use the $\theta = 0.18$ case of \citet{dbkn+06}. $\theta$
is the ratio of nova rates per unit $r'$ flux of the disc and bulge population.}
for M31. Our results are in good agreement with this number.
In order to make a further comparison, we compare the distributions of predicted and observed 
mass loss time distribution of novae in the following section.
 
In order to understand the evolution of nova populations with time, in
Fig.~\ref{fig:contour_diff_times}, we show the isodensity contours of
nova parameters at different stellar ages for elliptical-like galaxies
in our a025 model. Using \citet{ypsk05} data we can find the maximum luminosity
of novae and then convert it to $M_{\rm v}$ (see the following
section). From these plots, it is found that the average WD mass of a
nova population decreases with increasing stellar ages, which is
consistent with previous studies, e.g. \citet{poli96}. According to
the results of \citet{ypsk05}, the mass loss times and recurrence
periods of novae with massive WDs are shorter compared with novae with
smaller WD mass \citep[see also][]{tl86,pltw90}. Therefore, it is expected that the typical mass loss
time and recurrence time of populations of novae will increase with increasing stellar age. Similarly, the typical maximum magnitude of
novae become lager as the stellar age increases.

\subsection{Current nova population}
\label{subsec:cur_nova}

\begin{figure*}   
	\includegraphics[width=\textwidth]{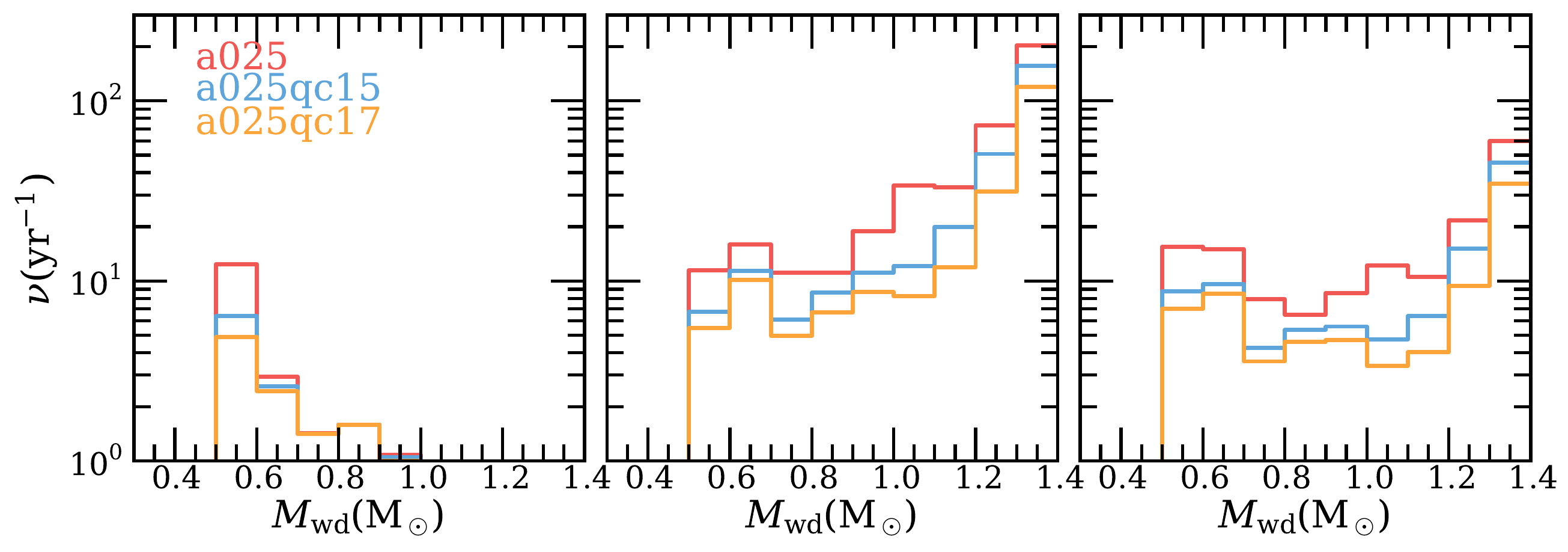}
    \caption{Distribution of nova rate as a function of WD mass for current nova population of
      elliptical-like galaxies (left panel), spiral-like galaxies (middle panel)
      and M31-like galaxies (right panel) in different models (see table~\ref{tab:model_tab}). 
      The red, blue, orange colours are for a025, a025qc15, a025qc17 model, respectively. 
    }
    \label{fig:mwd_hist}
\end{figure*}

\begin{figure*}   
	\includegraphics[width=\textwidth]{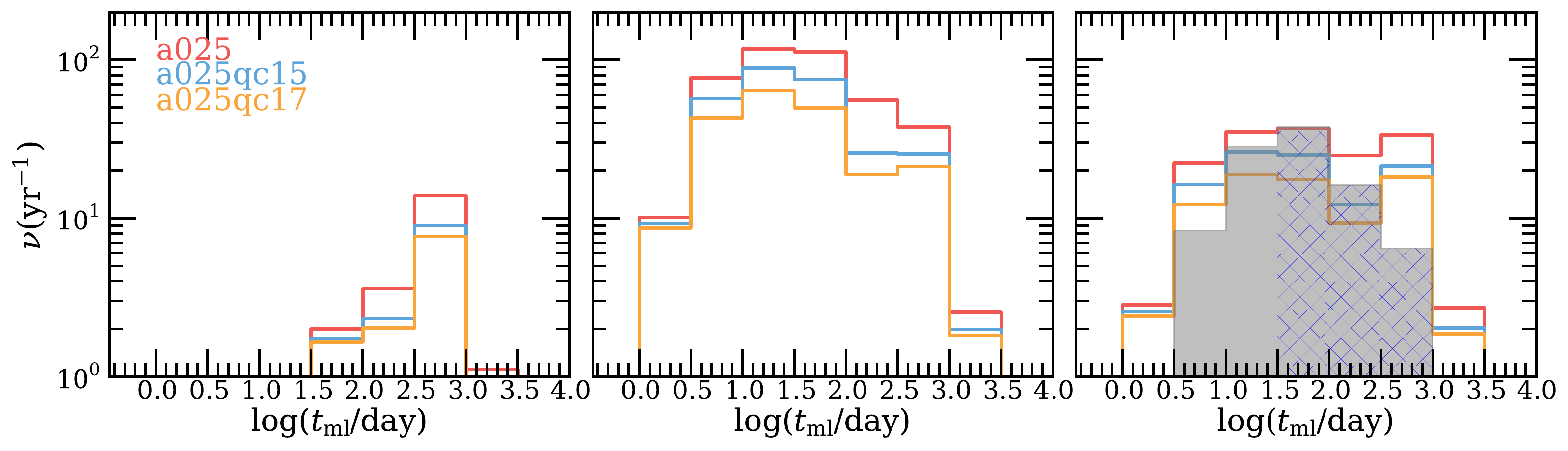}
    \caption{Mass loss time distribution of current nova population 
      of elliptical-like galaxies (left panel), spiral-like galaxies (middle panel) 
       and M31-like galaxies (right panel) in different models (see table~\ref{tab:model_tab}).
     The red, blue, orange colours are for a025, a025qc15, a025qc17 model, respectively. 
     The gray histogram shows the combined observational nova data from \citet{arp56} and \citet{dbkn+06}
     taking the incompleteness into consideration \citep{sg14,sgwb15}.
     The shaded histogram shows the observational nova data of Darnley's paper only.}
    \label{fig:tml_hist}
\end{figure*}

\begin{figure*}  
	\includegraphics[width=\textwidth]{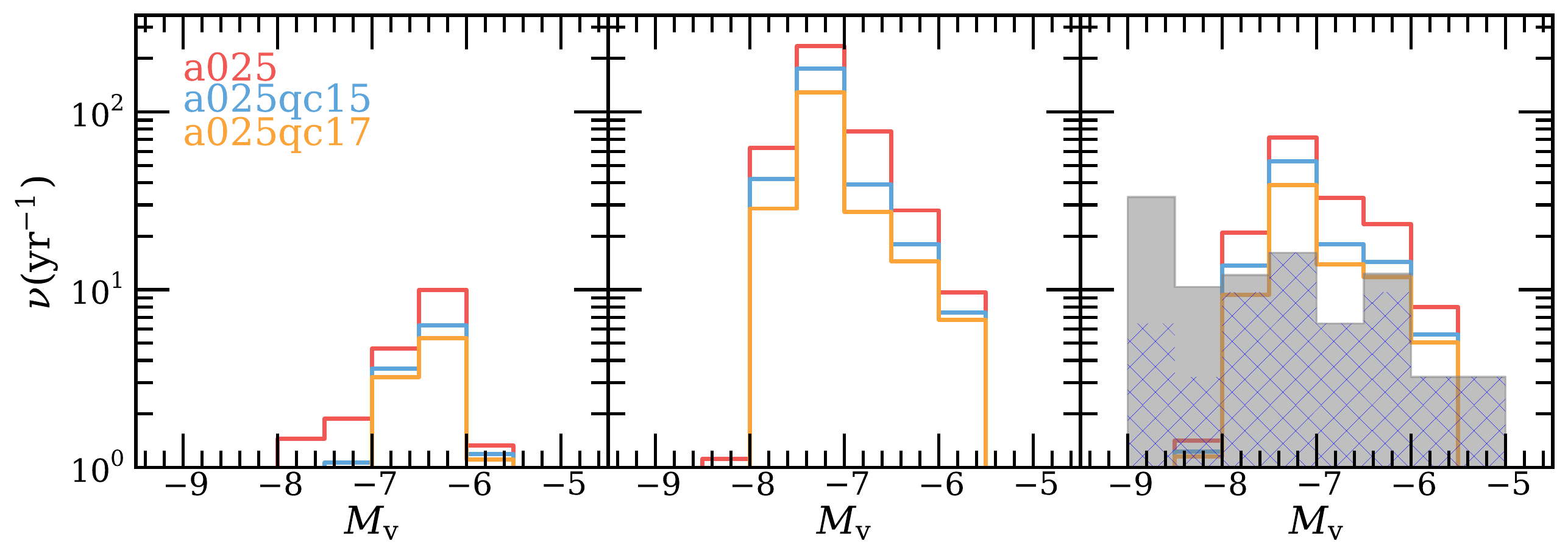}
    \caption{Distribution of V band maximum magnitude for current nova 
    population of elliptical-like galaxies (left panel), spiral-like galaxies (middle panel) and 
    M31-like galaxies (right line) in different models (see table~\ref{tab:model_tab}). The red, 
    blue, orange colours are for a025, a025qc15, a025qc17 model, respectively. The gray histogram shows the combined 
    observational nova data from \citet{arp56} and \citet{dbkn+06} taking the incompleteness
    into consideration \citep{sg14,sgwb15}. The shaded histogram shows the observational nova data 
    from \citet{dbkn+06} only.
      }
    \label{fig:mv_hist}
\end{figure*}

\begin{figure*}  
	\includegraphics[width=\textwidth]{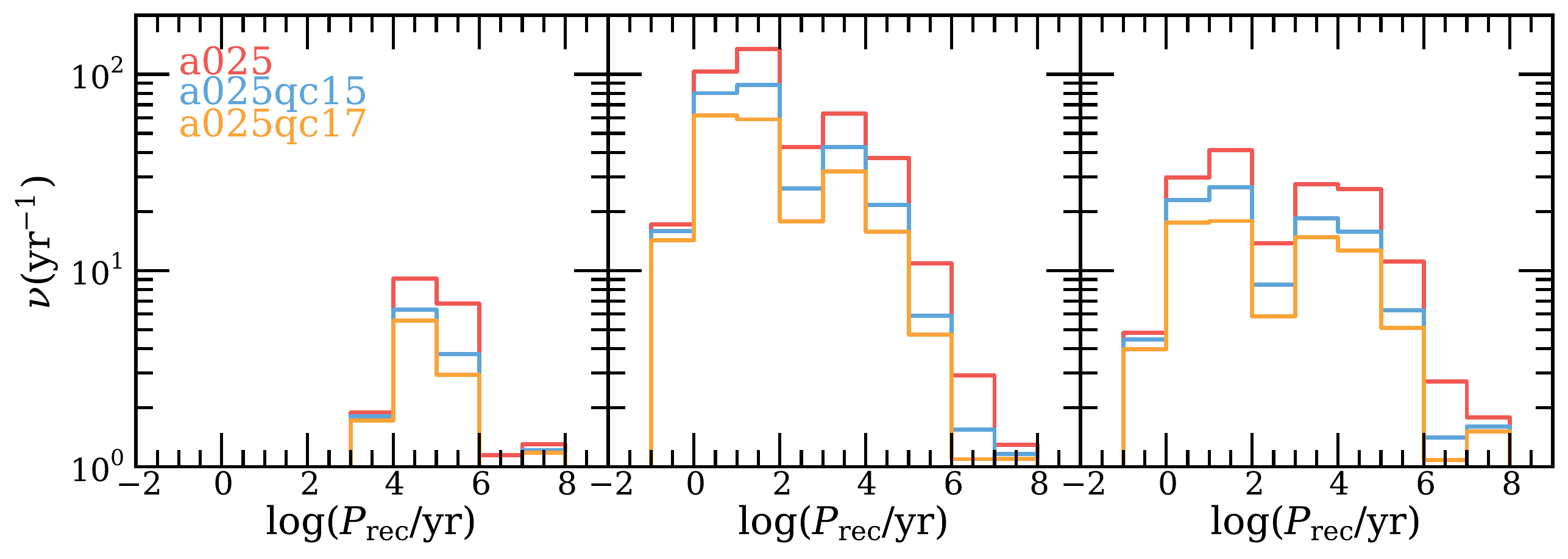}
    \caption{ Distribution of recurrence period for current nova population
      of elliptical-like galaxies (left panel), spiral-like galaxies (middle panel)
      and M31-like galaxies (right panel) in different models (see table~\ref{tab:model_tab}). The 
      red, blue, orange colours are for a025, a025qc15, a025qc17 model, respectively. }
    \label{fig:prec_hist}
\end{figure*}

Since the properties of novae are
mainly determined by the WD mass, we show the WD mass spectra of the
``current'' (10 Gyr) nova population in Fig.~\ref{fig:mwd_hist}. The WD mass peaks
around $0.5-0.6\,\rm M_{\odot}$ in elliptical-like galaxies and
$1.30-1.40\,\rm M_{\odot}$ in spiral-like galaxies. This is expected;
in elliptical-like galaxies, WD binaries with massive WDs evolve
faster and the number of massive WDs decreases. Therefore, the typical
WD mass of novae at 10 Gyr is small. The cut off at the low mass side is
simply due to the fact that we ignore He WDs in our calculation. In spiral-like
galaxies, there are young stellar populations with massive WDs. 
Most of these WDs have initially small mass and have increased their mass
during the thermal timescale mass transfer (see Fig. 2 of \citet{cwyg+14}). From
\citet{ypsk05}, we know that the ignition mass of novae for massive WDs
can be smaller by $2-3$ orders of magnitude, compared with low mass
WDs. Therefore, massive WDs can have relatively more frequent nova
outbursts. In M31-like galaxies, there are two peaks in the WD
spectra. The peak around low WD mass is simply due to the large number
of WD binary systems with low mass WDs. The peak around massive WD
mass is due to massive WDs having more frequent nova outbursts.

In Fig.~\ref{fig:tml_hist}, we show the mass loss time ($t_{\rm ml}$)
distribution of novae for different kinds of galaxies and compare the
result of M31-like galaxies with the observational data of M31. With
respect to the observational data, identical to the above description of
the observed nova rate in M31, we combine the nova data of \citet{arp56}
and \citet{dbkn+06}.  With the results of \citet{ypsk05}, we can get
two timescales for each novae - the decay time of bolometric luminosity
by 3 mag $t_{\rm 3,bol}$ and the duration of mass loss phase $t_{\rm
  ml}$. It is found that $t_{\rm ml}$ is much closer to the observed
$t_{3}$ than $t_{\rm 3,bol}$ \citep{pk95,ypsk05}.  The following
formula is used to compute the values of $t_{\rm ml}$ for observed
novae \citep{be08}.\\
$
t_{2} =
       \begin{cases}
       t_{\rm ml}/2.1  & t_{\rm ml} < 50\; {\rm days}  \\
       t_{\rm ml}/1.75 & t_{\rm ml} \geq 50\; \mbox{days}
       \end{cases}
$\\       
One point worth noting is that using a higher and fixed critical mass
ratio in the CE criteria does not result in an unrealistic nova population. 
In elliptical-like galaxies, novae have a typical mass loss time
around hundreds of days. The mass loss time of novae in spiral-like
galaxies peaks around several tens of days. In M31-like galaxies, the
mass loss time of novae ranges from tens of days to hundreds of
days. Compared with the observational data of M31, there are
too many novae predicted with $t_{\rm ml} < 10$ days and $t_{\rm ml} >
300$ days. Although \citet{sg14} and \citet{sgwb15} 
have corrected for the incompleteness of the short events, their correction was approximate 
and may not be accurate enough for the shortest novae. No incompleteness
was applied for the longest events in our comparison.
The novae with the shortest mass loss times are easily missed in observations, 
since observations are usually discontinuous. On the
other hand, the novae with the longest mass loss times may be difficult to 
detect, because these novae are likely to be faint and the
luminosity may not dramatically decline within the limited observational
time.

Following \citet{srqk+09}, we assume that the colour of novae at
maximum luminosity is the same as the colour of an A5V star ($T_{\rm eff}
\sim 8200\,{\rm K}$). Then we can get the V band magnitude of novae
for any given maximum bolometric luminosity with the data compiled by
\citet{john66}. In Fig.~\ref{fig:mv_hist}, we show the distribution of
V-band magnitude for novae at maximum luminosity. The novae are
dominated by those with $M_{\rm v}$ from about -7.0 to about -6.0 in elliptical-like
galaxies while absolute V-band magnitude peaks around about -8.0 to -6.5 in
spiral-like and M31-like galaxies. \citet{arp56} found that the
maximum photographic magnitude of novae in M31 has a bimodal
distribution. However, in M31-like galaxies, we do not find a bimodal
distribution in the theoretical and observational results.  The
observational data of M31 is shown in the right panel of
Fig.~\ref{fig:mv_hist}, which is based on the results of \citet{arp56}
and \citet{dbkn+06}. In the Arp's survey, the
photographic magnitudes were given.  We correct them for the foreground
extinction with $A_{\rm pg} = 0.25$ \citep{srqk+09}.  Then we convert
the photographic magnitude to V-band magnitude using colours $(B-m_{\rm
  pg}) = 0.17$ \citep{arp56,cdrd89} and $(B-V) = 0.15$ \citep{srqk+09}.  For
novae from \citet{dbkn+06}, we convert the observed magnitude to
V-band magnitude using colour $(V-R) = 0.16$ \citep{srqk+09}.  Compared
with the observational data, we underestimate the number of
very bright novae. The possible reasons for this discrepancy
are as follows. First, we did not include low temperature WDs 
($T_{\rm c} < 1.0\times 10^{7}$\;K) in our
calculation. With the same code used by \citet{ypsk05}, \citet{sypk+10}
(see also \citet{skp94}) found that some very luminous novae can be
explained by low temperature WDs with low accretion rates. 
Additionally, as \citet{pk95} discussed, the error in the maximum luminosity
in their simulations can be as large as $-0.75$ magnitude.
Finally, in our calculation of $M_{\rm v}$, we
  assume that the spectra of all novae at maximum are similar to normal
  A5V main sequence stars. However, previous studies
  \citep[e.g.][]{be08} found that the spectra of novae at maximum
  resemble the spectra of stars with spectral type in the range B5 to
  F5. Recently, based on high quality photometric spectra of novae,
  \citet{muna14} found that the spectra of some novae deviate from A5
  main sequence stars (see his Fig. 4). For example, the colour (B-V)
  ranges from around 0.20 to 3.40, while (B-V) = 0.15 for A5V main sequence stars. 
  This evidence suggests that adopting a unique spectral type may 
  not be an accurate assumption. This will influence the colour
  and bolometric correction used in the above conversion.

In Fig.~\ref{fig:prec_hist}, we show the distribution of recurrence periods
for novae in different galaxy types given different models. The novae in 
elliptical-like galaxies have relatively long recurrence periods, while 
the novae in spiral-like galaxies have predominantly shorter recurrence periods.
In M31-like galaxies, there are two peaks, which correspond to 
the two peaks of the WD mass distribution in Fig.~\ref{fig:mwd_hist}.
The peak around massive WD produces the peak around short recurrence 
periods; most of these WDs have accumulated additional mass during a prior thermal 
timescale mass transfer phase. These massive WDs have more frequent outbursts and 
are also short lived (see Fig.~\ref{fig:contour_diff_times}). The peak around small 
WD masses corresponds to the peak around
long recurrence periods. In spiral-like galaxies, there is a large number 
of WD binaries with small WD mass. But these small WDs have less 
frequent outbursts. Given the short observational time  in
reality, less novae, particularly recurrent novae, will be detected in
elliptical galaxies. If we take the novae with recurrence period
$P_{\rm rec} < 100$\, yr as recurrent novae, there will be no
recurrent novae in elliptical-like galaxies. In spiral-like galaxies,
the fraction of recurrent novae is around $\sim 60\% - 65\%$. In M31-like galaxies, the fraction of recurrent novae
is $\sim 45\% -
50\%$. \citet{shrs+15} performed a thorough astrometric study of novae in
M31 and did a Monte Carlo analysis of the detection effficiency of recurrent novae.
They suggested that as many as one in three M31 novae may be
recurrent novae with $P_{\rm rec} < 100$\,yr. Our result is not inconsistent 
with their results.

\section{Discussion}
\label{sec:disc}

\subsection{Influence of $\alpha$ values}

\begin{figure}   
	\includegraphics[width=\columnwidth]{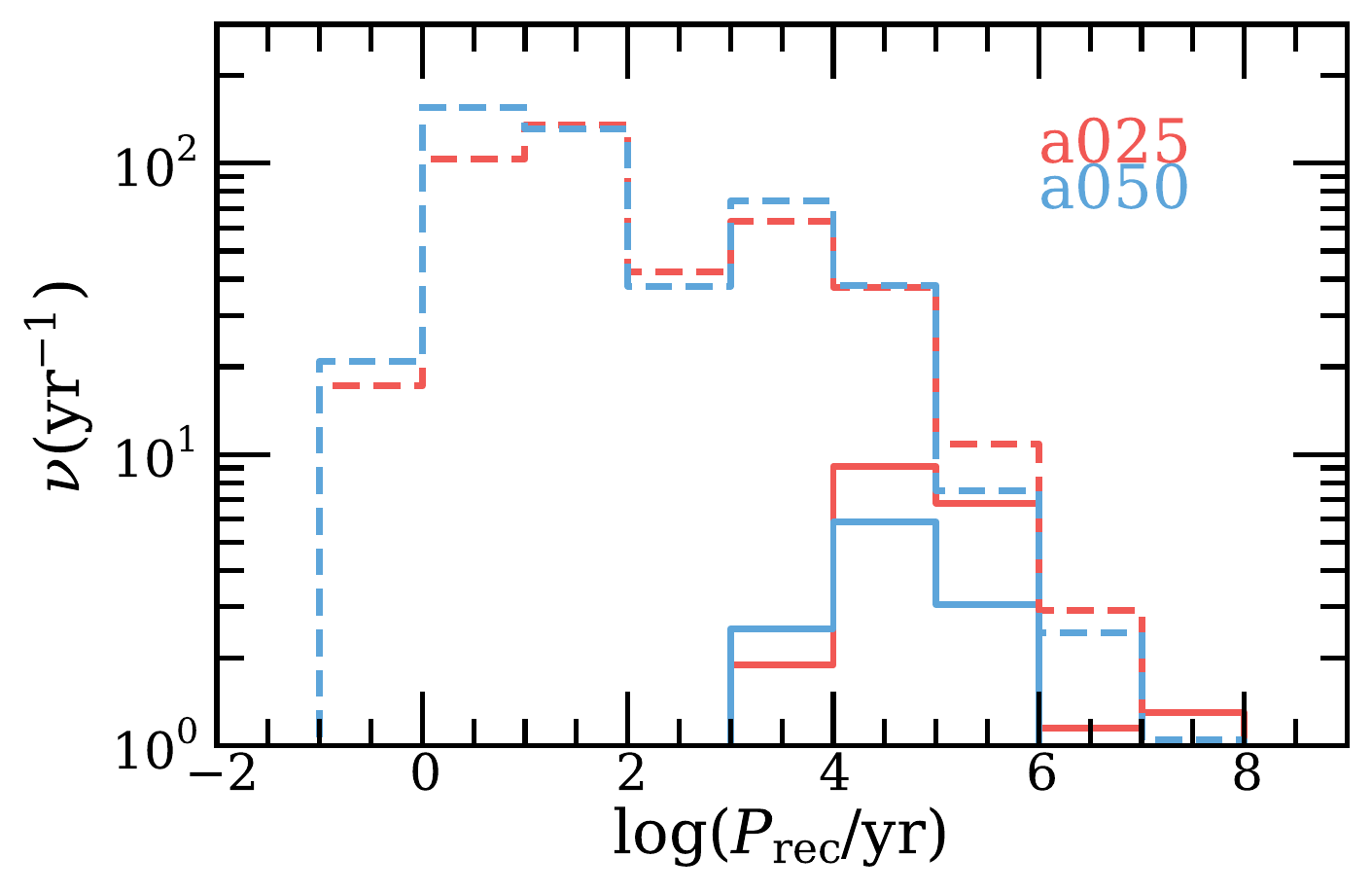}
    \caption{Comparison of recurrence period distribution of current nova population
      for elliptical-like (solid line) and spiral-like galaxies (dashed line) in a025
      model (red line) and a050 model (blue line).  }
    \label{fig:prec_hist_diff_al}
\end{figure}

As we discussed in section~\ref{sec:bps}, the appropriate value(s) of $\alpha$
suffers from considerable uncertainty.  Therefore, we computed a model with
$\alpha = 0.5$ (a050 model) and found no dramatic
difference between the results of this model and a025 model, which
is similar to what we found in \citet{cwyg+14}. Here we only show the
distribution of recurrence period for comparison in Fig.~\ref{fig:prec_hist_diff_al}. 
The influence of $\alpha$ values on the evolution of the binary population is 
rather complicated. Qualitatively it is clear that for larger $\alpha$ values more 
binaries will survive from the first CE phase and typical binary separations of 
survivors will become larger. For larger binary separations, in more binary systems
the mass loss of nondegenerate donors upon ROLF will be unstable and a fraction 
of population underlaying novae will be lost. As well, in remaining systems accretion 
rates will be higher and WD may spend more time as stably nuclear burning objects
and less as nova progenitors (less mass will be left for unstable burning). But 
the last effect may be offset by the increased WD mass and WDs with higher accretion
rates have lower ignition mass. As a result,the total nova rates remain 
comparable. A more thorough study of this issue is beyond the scope of the present 
paper.

\subsection{Influence of WD interior temperatures}

In the above nova calculation, we assume that the interior temperatures of WDs are constant,
i.e. $T_{\rm c} = 1.0\times10^{7}$\,K. In order to understand the influence of 
WD temperature on our results, we computed a model with $T_{\rm c} = 3.0\times10^{7}$\;K.
The other assumptions of this model are the same as in a025 model.

\begin{figure}  
	\includegraphics[width=\columnwidth]{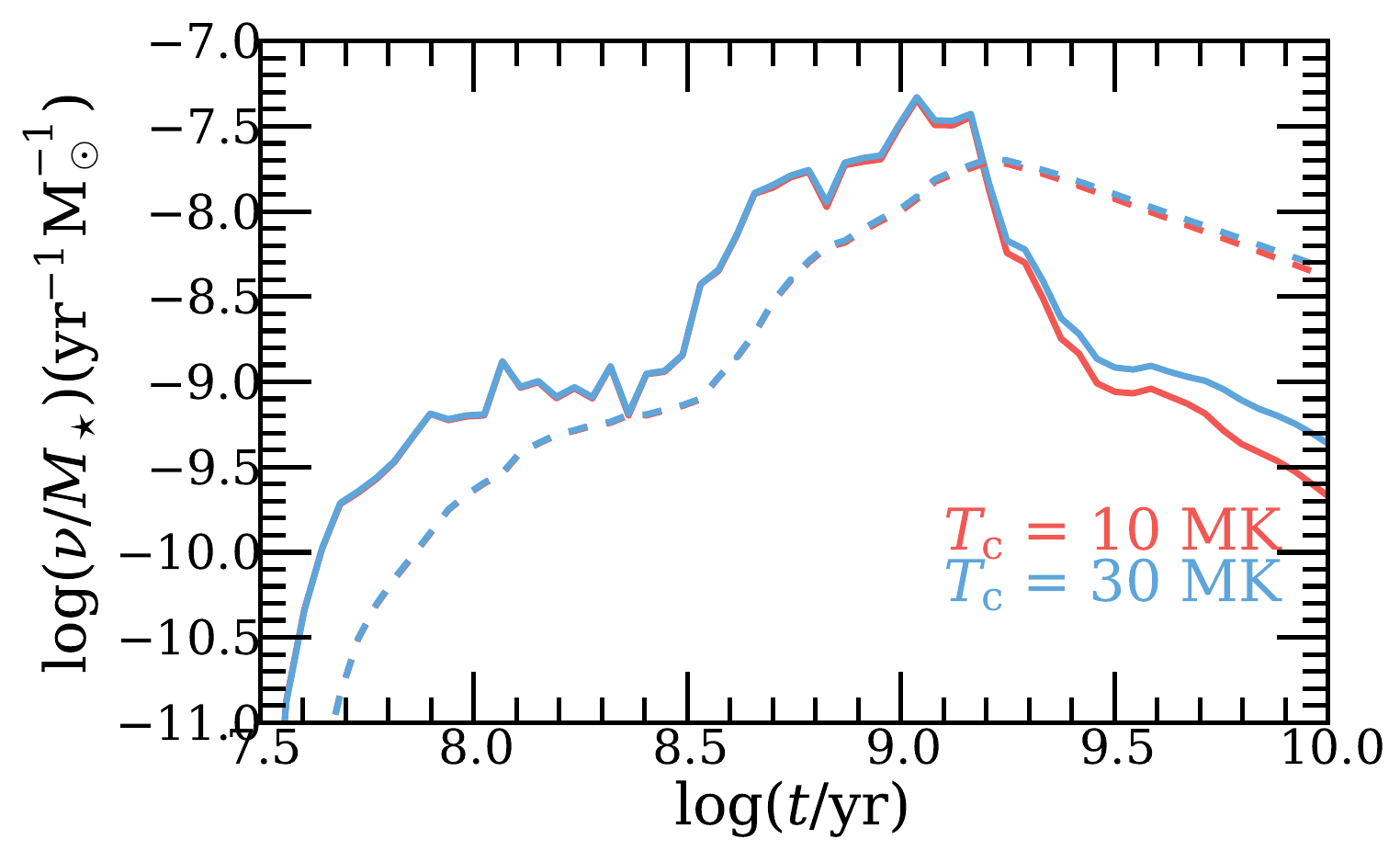}
    \caption{Mass-normalized nova rates as a function of stellar age for elliptical-like galaxies
    (solid line) and spiral-like galaxies (dashed line) in a025 model assuming WD temperatures
     $T_{\rm c} = 1\times 10^{7}$ K (red colour), $T_{\rm c} = 3\times 10^{7}$ K (blue colour).}
    \label{fig:rnova_diff_twd}
\end{figure}

\begin{figure}  
	\includegraphics[width=\columnwidth]{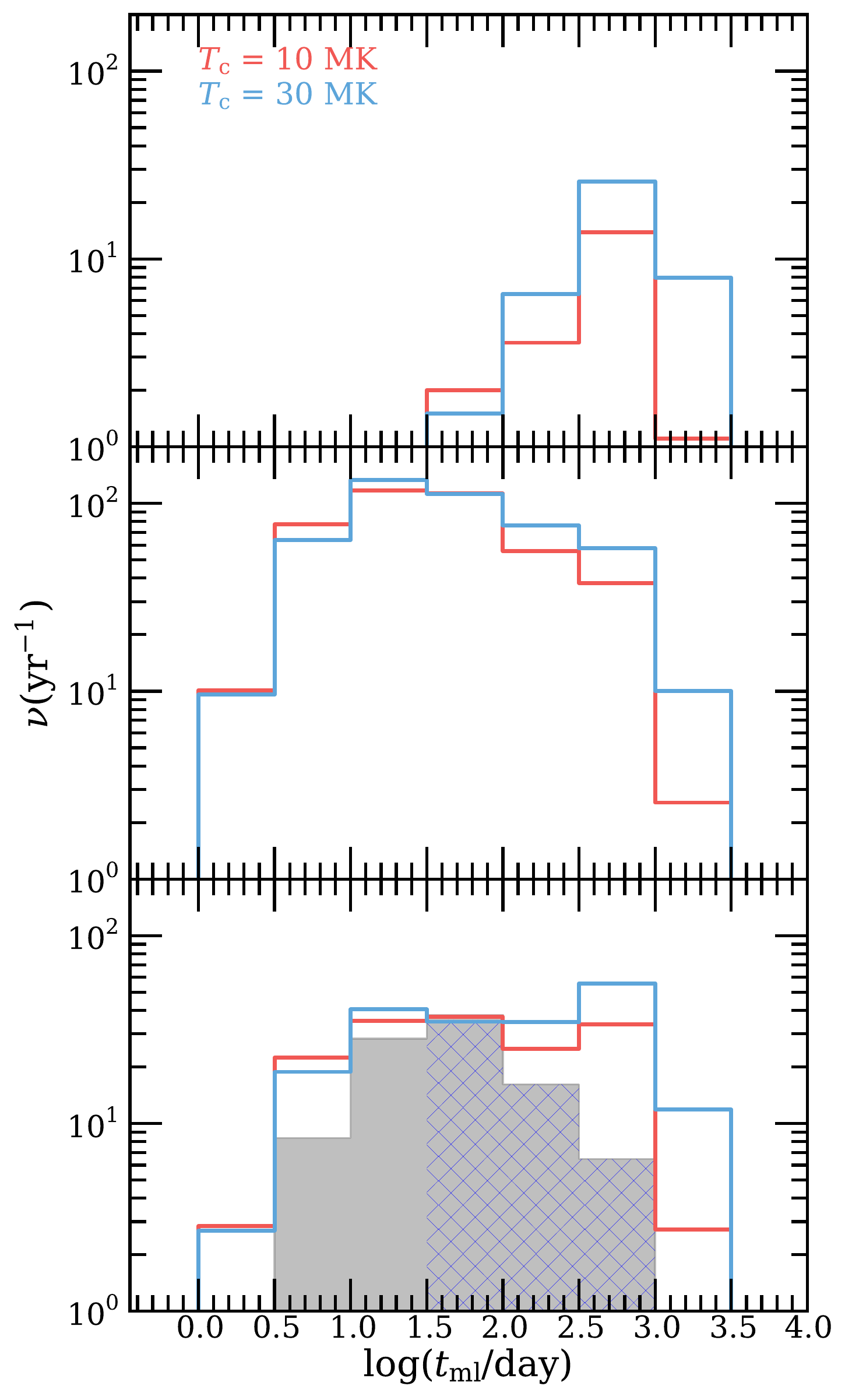}
    \caption{Mass loss time distribution of current nova population 
    for elliptical-like galaxies (upper panel), spiral-like galaxies (middle panel) 
    and M31-like galaxies (lower panel) in a025 model assuming WD temperatures
     $T_{\rm c} = 1\times 10^{7}$ K (red colour), $T_{\rm c} = 3\times 10^{7}$ K (blue colour).
     The gray histogram shows the observational
     nova data from \citet{arp56} and \citet{dbkn+06} taking the incompleteness into 
     consideration \citep{sg14,sgwb15}. The shaded histogram shows the observational 
     nova data of \citet{dbkn+06} only.}
    \label{fig:tml_hist_diff_twd}
\end{figure}

\begin{figure}  
	\includegraphics[width=\columnwidth]{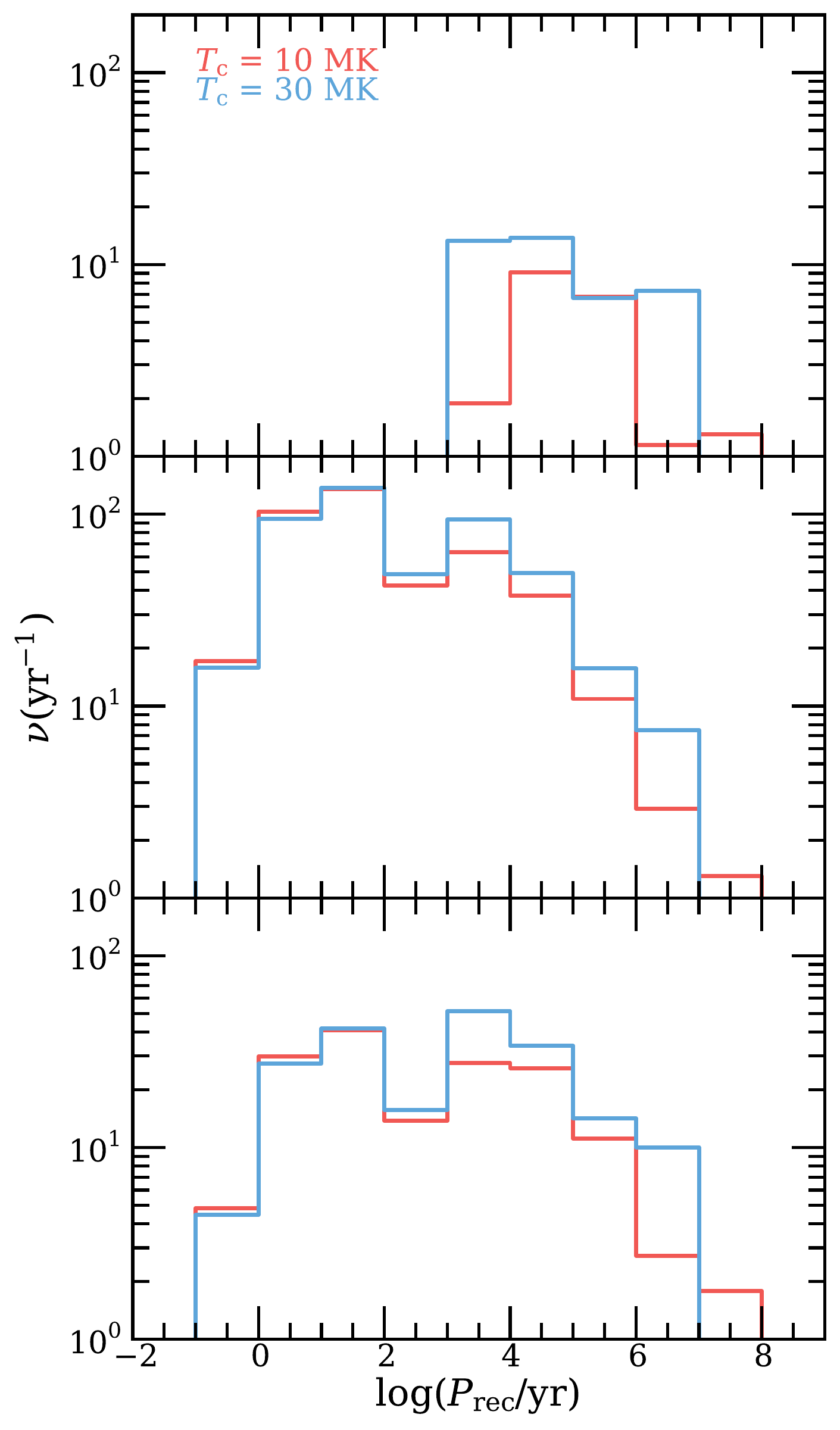}
    \caption{Recurrence period distribution of current nova population 
     for elliptical-like galaxies (upper panel), spiral-like galaxies (middle panel) and M31-like 
    galaxies (lower panel) in a025 model assuming WD temperatures $T_{\rm c} = 1\times 10^{7}$ K (red colour), $T_{\rm c} 
= 3\times 10^{7}$ K (blue colour). }
    \label{fig:prec_hist_diff_twd}
\end{figure}

From Fig. \ref{fig:rnova_diff_twd}, we see that there is no
dramatic difference among models with different WD temperatures and
nova rates are only slightly higher for WDs with higher temperatures (see also table~\ref{tab:model_tab}).
This can be understood since the ignition mass of novae is lower for WDs
with higher temperature \citep{ypsk05}. In elliptical-like galaxies,
the difference becomes larger at old ages. This is due to the fact that the difference
in ignition mass at different temperatures is larger for low WD masses than for massive WD \citep{ypsk05}.

In Fig.~\ref{fig:tml_hist_diff_twd}, we show
the distribution of mass loss time at 10 Gyr in different types of galaxies for 
a025 model assuming different WD temperatures. Compared with model
with WD temperature $T_{\rm c} = 1.0 \times 10^{7}$ K, there is an
enhancement of novae with long mass loss times ($t_{\rm ml} > 100$ days)
for all galaxies types in the model with a higher WD temperature. 
In Fig.~\ref{fig:prec_hist_diff_twd}, we show the distribution of recurrence period of novae.
There is an enhancement of novae with long recurrence period for the model with a high WD temperature. 
The trend seen in Figs.~\ref{fig:tml_hist_diff_twd} and \ref{fig:prec_hist_diff_twd} 
is explained by the fact that, at $\dot{M}_{\rm acc} \leq 10^{-8}\,M_{\odot}\,{\rm yr}^{-1}$
at which most of novae occur, $P_{\rm rec}$ drops with increase of $T_{\rm c}$, while 
$t_{\rm ml}$ increases with increase of $T_{\rm c}$ \citep{ypsk05}.

\subsection{Influence of metallicity}

\citet{sfbm+10} found that in the inner few arcsecs of the bulge of M31, the 
metallicity is $\sim 3\,Z_{\odot}$. Except for this region, the metallicity is solar. In addition, the metallicity of the disc
is less than solar. In our computation, we assume that the metallicity of the stellar 
population is solar. The influence of metallicity on our results is rather complicated.
First, it will influence the binary evolution. For example, \citet{mch08} found that,
for $Z\ge 2Z_{\odot}$, the final WD mass will increase as metallicity increases, while for metallicity 
$Z < 2Z_{\odot}$, it will decrease as metallicity increases. 
Additionally, \citet{pcit00} computed the evolution of accreting WDs for H-rich material with different
metallicities and they found that the accretion rate for stably burning WD will be lower for lower metallicity
and the width of the stably burning regime will be reduced. On the other hand, metallicity will influence 
the nova properties. For example, \citet{ssts00} have studied the effects of metallicity on nova outburst and 
found that the ignition mass is higher for lower metallicity. Although 
metallicity is important in our calculation, it is difficult for us to quantitatively assess its influence on 
our results, and further studies are needed in this regard.

\subsection{Novae with donors at differing evolutionary states}

\begin{figure}       
    \includegraphics[width=\columnwidth]{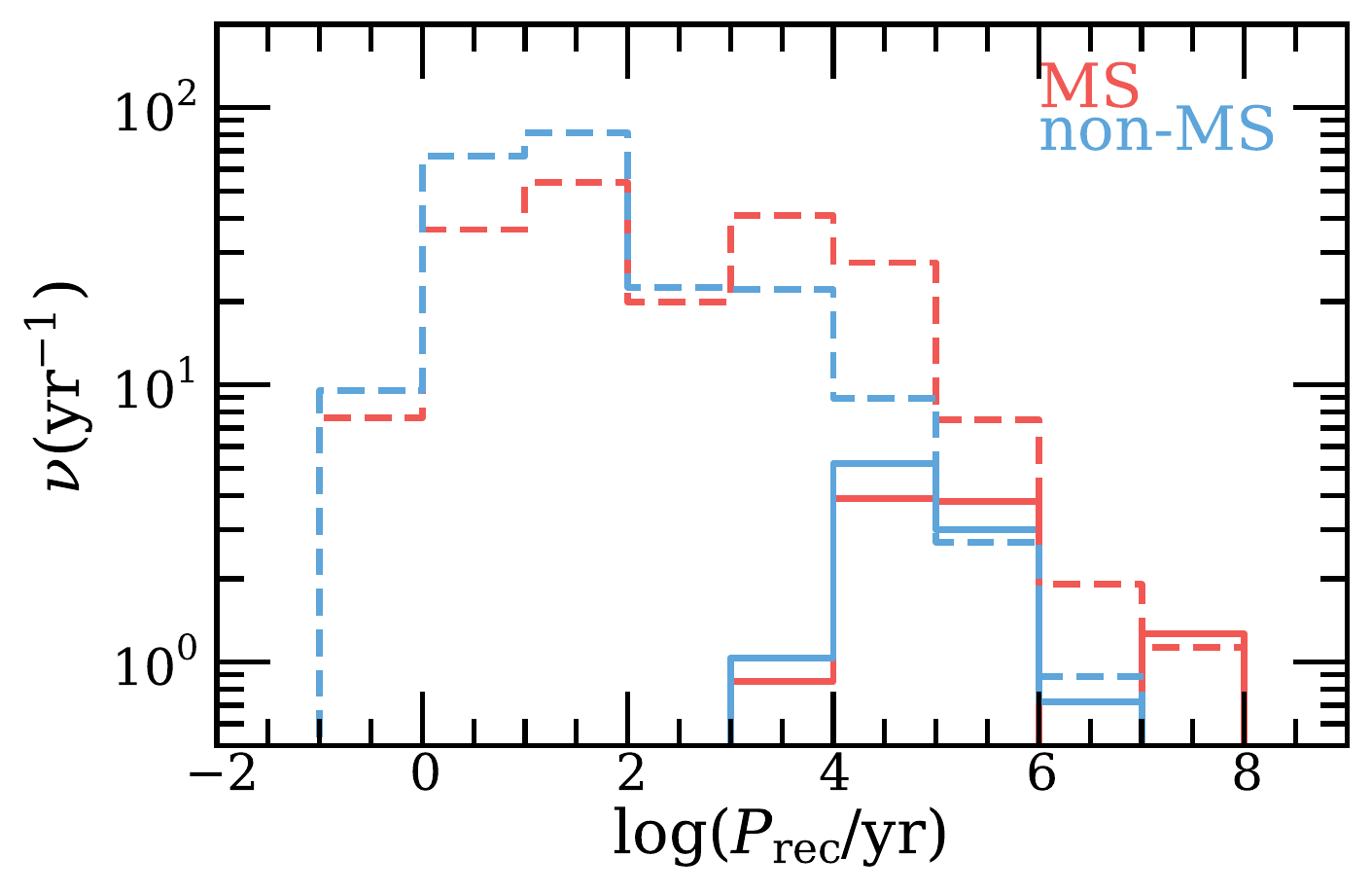}
    \caption{Distribution of recurrence period for current nova population 
with different types of donors in elliptical-like (solid line) and spiral-like galaxies (dashed line). The red and blue 
lines shows the novae with MS donors and non-MS donors (i.e. HGs and RGs), respectively. The donor type 
 is defined according to the donor type at the onset of mass transfer.} 
    \label{fig:prec_hist_diff_type}
\end{figure}

\begin{figure}       
    \includegraphics[width=\columnwidth]{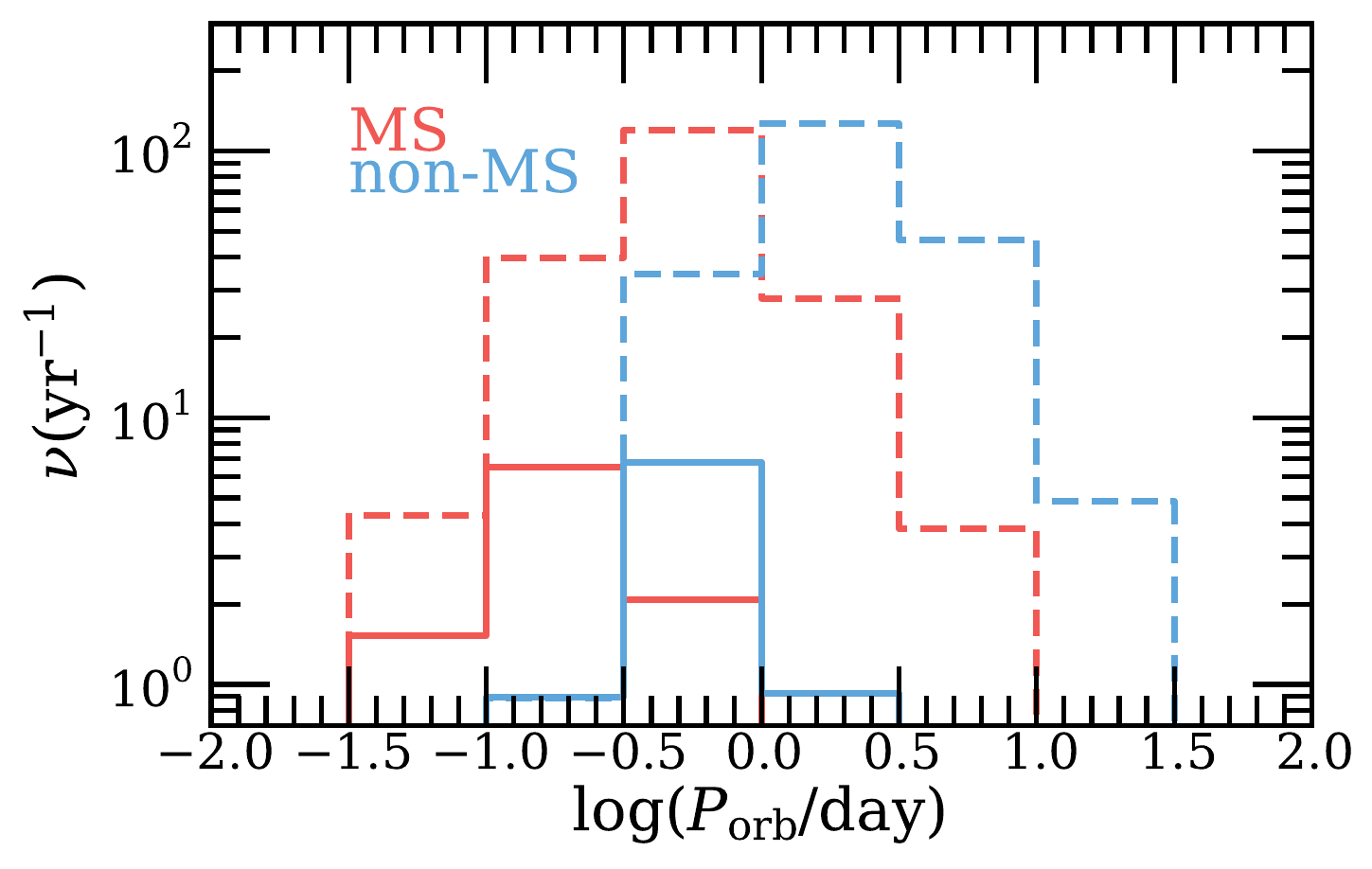}
    \caption{Distribution of orbital periods for current nova population 
    with different types of donor in elliptical-like (solid line) and spiral-like galaxies (dashed line). 
    The red and blue lines shows the novae with MS donors and non-MS donors (i.e. HGs and RGs), respectively.
    The donor type is defined according to the donor type at the onset of mass transfer.}
    \label{fig:porb_hist_diff_type}
\end{figure}

We classify novae according to the donor type at the onset of mass 
transfer. In Fig.~\ref{fig:prec_hist_diff_type}, we show the recurrence period distribution of the ``current'' nova 
population with different donor types in elliptical-like galaxies and spiral-like 
galaxies. The contribution to the total nova rate from WD binaries with non-MS donors is comparable to that of WD binaries with MS donors. This conclusion is also true for our model of a M31-like galaxy.
In Fig.~\ref{fig:porb_hist_diff_type}, we show the orbital period distribution of the current nova population
in elliptical-like and spiral-like galaxies. It is found that the secondaries of those novae with non-MS donors are mainly located in the hydrogen-shell burning stage.
Those WD binaries with evolved secondaries have relatively higher mass transfer rates. 
In addition, the WD will accumulate mass during the thermal timescale mass transfer. For higher mass transfer rates and large WD masses, the ignition mass will be smaller.
Therefore, there will be more frequent outbursts for these binaries.

\citet{wdbk+14} found 11 nova progenitor systems with evolved secondaries in a survey with 38 confirmed nova progenitor systems.
This leads to a fraction of $\sim 30\%$ of the nova progenitor systems which harbour evolved secondaries.
Given the nova progenitor systems with evolved secondaries have relatively higher mass transfer rates
compared with nova progenitor systems with MS donors, nova progenitor systems with evolved secondaries 
should have more frequent outbursts. Therefore, the contribution of novae with evolved donors to 
the total nova rate should be larger than the fraction of corresponding binary systems, $30\%$. 
Recently, \citet{wdbs16} conducted a more accurate statistical analysis of their preceding results and 
came to the conclusion that $\sim30^{+13}_{-10}\%$ of nova eruptions in M31 Galaxy are RG-novae. 

\subsection{Correlation between mass-specific nova rate and morphological type of galaxy}

Observations give contradicting results regarding the dependence of the mass-specific nova rate on the Hubble type of
the galaxy. Some observational studies \citep[e.g.][]{ctjf+90,scp00,fcj03,ws04} 
claimed that there is no strong dependence of mass-specific nova 
rate on the Hubble types of galaxies: the majority of galaxies have around   
$2 \pm 1\times10^{-10} {L_{\odot,K}}^{-1}\,{\rm yr}^{-1}$. However, 
there are several galaxies which may have a higher mass-specific nova rate,
e.g. the SMC, LMC, M33 and others\citep{drbl94,dell02,ns05,as14,sdlz+16}.  
The mass-specific nova rate of elliptical-like  
 galaxies at 10 Gyr in our calculation is around $(1-2) \times 10^{-10}\,{\rm M_{\odot}}^{-1}\,{\rm yr}^{-1}$. 
This is consistent with \citet{ws04}, since the total stellar mass-to-light
ratio $M_{\star}/L_{\rm K} \sim 1.0$, slightly depending on the colour of the galaxy \citep{bd01}. 
However, for spiral-like galaxies, we obtained a mass-specific nova rate  
10-20 times larger. It would seem then that our results are inconsistent
with measurements of \citet{ws04}. However, one crucial point to understand this discrepancy is 
the incompleteness of nova surveys. For example, the nova 
rate of M31 in the analysis of \citet{ws04} is significantly lower than the nova rates of \citet{dbkn+06}.
As we discussed above, a more reliable value should be $\approx97\,{\rm yr}^{-1}$, giving
a luminosity-specific nova rate $\sim 7.0\times10^{-10}\,{L_{\odot,K}}^{-1}\,{\rm yr}^{-1}$, which is a factor of $3-4$
larger than the value in \citet{ws04}. In our model for M31, we find a mass-specific nova rate of $\approx (7-14) \times 10^{-10}\,{\rm M_{\odot}}^{-1}\,{\rm yr}^{-
1}$, which is consistent with the above number.
In our model of spiral-like galaxies, we assume the SFR to be constant, which is almost certainly oversimplified in the general case.
As shown in Figs.~\ref{fig:rnova} and~\ref{fig:contour_diff_times}, if there was an initial spike 
of SFR, certain fraction of total mass of "realistic" spirals sits in old population
which does not contribute much to the current nova rate. This will lead to a smaller value of luminosity-specific
nova rate (cf. table~\ref{tab:model_tab}). 

\subsection{Novae with short recurrence periods}

In M31-like galaxies, the predicted rate of novae with recurrence period $P_{\rm rec} < 1$\,yr is 
around $4\,{\rm yr}^{-1}$ (see Fig.~\ref{fig:prec_hist}). The typical WD mass of these novae is $1.
30-1.40\,\rm M_{\odot}$ and their typical accretion 
rate is $10^{-7}-10^{-6}\,\rm M_{\odot}\,{\rm yr}^{-1}$. Recently, \citet{tbwl+14} discovered a 
nova with recurrence period of around 1 yr, and \citet{hdkn+15} found the recurrence period of this nova
is more likely to be 6 months. Based on the recurrence period and the X-ray emission, 
\citet{tbwl+14} constrained the WD mass of the 
nova to be $> 1.30\,\rm M_{\odot}$ and its accretion rate to be $>1.7 \times 10^{-7}\,\rm M_{\odot}\,{\rm yr}^{-1}$.
Presently, this is the only nova with a recurrence period less than 1 yr detected in M31 giving a nova rate $\sim$$2\;{\rm yr}^{-1}$.
We predict a rate of around $4\;{\rm yr}^{-1}$. Given the uncertainties of observations and simulations,
our results for short recurrence periods are consistent with observations.

\subsection{Novae with ONe WDs}

In our calculation, we do not distinguish ONe WDs from CO WDs. Evidently, some massive 
WDs should be ONe WDs. \citet{jh98} found that ONe WDs have higher ignition masses (about a factor of 2), 
compared with CO WDs. This is due to the lower $^{12}$C abundance in 
the envelope which will reduce the reaction rate of the CNO cycle and less energy will be 
released at the same temperature. Therefore, the ONe WDs should have longer recurrence periods.
However, the difference of $t_{\rm ml}$ and $M_{\rm V}$ between ONe WD and CO WD with same mass has not been investigated.
Assuming that the ignition mass of novae for ONe WD is the same as for CO WD, we compute the nova rate of ONe WDs based on 
the WD type identification in \textsc{bse} code. We find that,
for the current nova population in elliptical-like and spiral-like galaxies, the contribution of ONe WDs to the total nova rate
is rather small (less than $10\%$).

\subsection{Novae in Globular Clusters}

Motivated by observations that the mass-specific number of low mass X-ray binaries is enhanced 
in globular clusters (GCs) \citep[e.g.][]{clar75}, it is also expected that a significant population of WD binaries could 
be formed through dynamical process in GCs \citep[e.g.][]{bail91}. Observations from the Chandra X-Ray 
Observatory and the Hubble Space Telescope have revealed dozens of cataclysmic variables in GCs \citep[e.g.][]{ghem01,hgec+05,
plhv+02}. Consequently, we will expect the nova rate to be enhanced in GCs. So far, there are five novae 
found in GCs: one in M80 \citep{dlkt+10}, one in a GC of M87 \citep{szbl+04}, two in the GCs of M31 \citep{sq07,hphd+13}, and
one in a GC of M84 \citep{cspn+15}. Based on the two novae discovered in the GCs of M31 \citep{hphd+13}, the estimated nova rate in the M31 GC system 
is $0.05\,{\rm yr}^{-1}\,{\rm GC}^{-1}$. \citet{cspn+15} found two novae in the GCs of M87 and M84 and estimated that 
novae are likely enhanced by at least an order of magnitude in GCs compared with the field.
This evidence indicates that dynamical interactions in GCs may be an important factor in the calculation of the nova rate in a galaxy.
Obviously, in our calculation, we do not take GCs into consideration. 

\section{Conclusions}
\label{sec:con}

With a hybrid binary population synthesis method, we have modelled the nova population for a starburst 
(i.e. elliptical-like galaxies) and a constant SFR model (i.e. spiral-like galaxies).
In addition, we have also provided a composite model as an analog of M31 (i.e. M31-like galaxies). 
We have computed the nova rates and the nova properties, such as their distributions of mass loss time, maximum magnitude, for
different stellar populations. We compare these results with
observational data of M31. Our main results are summarized as
follows.

1) We computed the nova rates as a function of stellar age in
elliptical-like and spiral-like galaxies (see
Fig.~\ref{fig:rnova}). In elliptical-like galaxies, the mass normalized nova rate peaks around 1\;Gyr
and declines by $\sim 2$ orders of magnitude at 10\;Gyr. In spiral-like galaxies, it peaks around 2\;Gyr
and declines by a factor of $\sim 4$ in spiral-like galaxies at 10\;Gyr. 
The mass-specific nova rate for elliptical-like galaxies at 10\,Gyr 
in our calculation is $\sim (1-2)\times 10^{-10}\,{\rm M_{\odot}}^{-1}\,{\rm yr}^{-1}$,
which is consistent with observations. However, the mass-specific nova rate for 
spiral-like galaxies at 10\,Gyr is $\sim (20-40)\times 10^{-10}\,{\rm M_{\odot}}^{-1}\,{\rm yr}^{-1}$,
which is larger than seen in some observations. The mismatch may be due to both the incompleteness of past surveys 
and the assumption of a constant SFR in our model. Moreover, our predictions go in the direction
suggested by several papers which found high luminosity-specific nova rates in 
late-type systems like LMC and M33 \citep[e.g.][]{drbl94,dell02,as14}.

2) The current nova population is dominated by novae with low mass
WDs in elliptical-like galaxies and by novae with massive WDs in
spiral-like galaxies.

3) In elliptical-like galaxies, the majority of current novae
have long mass loss times, are relatively faint, and have long 
recurrence periods. In spiral-like galaxies, the
majority of the current nova population have short mass loss times,
are relatively bright, and have short recurrence periods.

4) Given the uncertainties in both our calculation and observations, the
predicted nova rate and the distribution of nova mass loss times in our
M31-like galaxy are in good agreement with observational data for M31. The observed distribution may be 
subject to incompleteness at $t_{\rm ml} < 10$ day and $t_{\rm ml} > 300$ day.
In addition, it is possible that we underestimate the number of
very bright novae in our calculations. This may be due to the lack of
low temperature WDs or the assumption of a unique spectral type for
novae at the maximum luminosity in our work.

\section*{Acknowledgements}

We would like thank the referee for useful comments, which
helped to improve the paper. 
HLC would like to thank Monika Soraisam and Hans Ritter for helpful discussion about
the calculation of novae. We are grateful to the \textsc{mesa} council
for the \textsc{mesa} instrument papers and website. HLC gratefully
acknowledges support and hospitality from the MPG-CAS Joint Doctoral
Promotion Program (DPP) and the Max Planck Institute for Astrophysics
(MPA). Z. H. is partially supported by the National Natural Science
Foundation of China (Grant No.11390374,11521303), Science and
Technology Innovation Talent Programme of Yunnan Province (Grant
No. 2013HA005) and the Chinese Academy of Sciences (Grant
Nos. XDB09010202, KJZD-EW-M06-01). The work was partially
supported by Basic Research Program P-7 of the Presidium of the Russian Academy of Sciences 
and RFBR grants No.~14-02-00604 and 15-02-04053. 
LRY gratefully acknowledges warm hospitality and support from MPA-Garching.
MG acknowledges hospitality of the Kazan Federal University (KFU) and support by the Russian Government Program of Competitive Growth of KFU.
HLC acknowledges the computing time granted by the Yunnan Observatories and provided on
the facilities at the Yunnan Observatories Supercomputing Platform.


\bibliographystyle{mnras}

\begin{thebibliography}{}
\makeatletter
\relax
\def\mn@urlcharsother{\let\do\@makeother \do\$\do\&\do\#\do\^\do\_\do\%\do\~}
\def\mn@doi{\begingroup\mn@urlcharsother \@ifnextchar [ {\mn@doi@}
  {\mn@doi@[]}}
\def\mn@doi@[#1]#2{\def\@tempa{#1}\ifx\@tempa\@empty \href
  {http://dx.doi.org/#2} {doi:#2}\else \href {http://dx.doi.org/#2} {#1}\fi
  \endgroup}
\def\mn@eprint#1#2{\mn@eprint@#1:#2::\@nil}
\def\mn@eprint@arXiv#1{\href {http://arxiv.org/abs/#1} {{\tt arXiv:#1}}}
\def\mn@eprint@dblp#1{\href {http://dblp.uni-trier.de/rec/bibtex/#1.xml}
  {dblp:#1}}
\def\mn@eprint@#1:#2:#3:#4\@nil{\def\@tempa {#1}\def\@tempb {#2}\def\@tempc
  {#3}\ifx \@tempc \@empty \let \@tempc \@tempb \let \@tempb \@tempa \fi \ifx
  \@tempb \@empty \def\@tempb {arXiv}\fi \@ifundefined
  {mn@eprint@\@tempb}{\@tempb:\@tempc}{\expandafter \expandafter \csname
  mn@eprint@\@tempb\endcsname \expandafter{\@tempc}}}

\bibitem[\protect\citeauthoryear{{Alis} \& {Saygac}}{{Alis} \&
  {Saygac}}{2014}]{as14}
{Alis} S.,  {Saygac} A.~T.,  2014, in {Woudt} P.~A.,  {Ribeiro} V.~A.~R.~M.,
  eds,  Astronomical Society of the Pacific Conference Series Vol. 490, Stellar
  Novae: Past and Future Decades. p.~95

\bibitem[\protect\citeauthoryear{{Arp}}{{Arp}}{1956}]{arp56}
{Arp} H.~C.,  1956, \mn@doi [\aj] {10.1086/107284}, \href
  {http://adsabs.harvard.edu/abs/1956AJ.....61...15A} {61, 15}

\bibitem[\protect\citeauthoryear{{Bailyn}}{{Bailyn}}{1991}]{bail91}
{Bailyn} C.~D.,  1991, in {Janes} K.,  ed.,  Astronomical Society of the
  Pacific Conference Series Vol. 13, The Formation and Evolution of Star
  Clusters. pp 307--323

\bibitem[\protect\citeauthoryear{{Barmby} et~al.,}{{Barmby}
  et~al.}{2007}]{babe+07}
{Barmby} P.,  et~al., 2007, \mn@doi [\apjl] {10.1086/511682}, \href
  {http://adsabs.harvard.edu/abs/2007ApJ...655L..61B} {655, L61}

\bibitem[\protect\citeauthoryear{{Bell} \& {de Jong}}{{Bell} \& {de
  Jong}}{2001}]{bd01}
{Bell} E.~F.,  {de Jong} R.~S.,  2001, \mn@doi [\apj] {10.1086/319728}, \href
  {http://adsabs.harvard.edu/abs/2001ApJ...550..212B} {550, 212}

\bibitem[\protect\citeauthoryear{{Bode} \& {Evans}}{{Bode} \&
  {Evans}}{2008}]{be08}
{Bode} M.~F.,  {Evans} A.,  2008, {Classical Novae}

\bibitem[\protect\citeauthoryear{{Capaccioli}, {Della Valle}, {Rosino}  \&
  {D'Onofrio}}{{Capaccioli} et~al.}{1989}]{cdrd89}
{Capaccioli} M.,  {Della Valle} M.,  {Rosino} L.,   {D'Onofrio} M.,  1989,
  \mn@doi [\aj] {10.1086/115104}, \href
  {http://adsabs.harvard.edu/abs/1989AJ.....97.1622C} {97, 1622}

\bibitem[\protect\citeauthoryear{{Chen} \& {Han}}{{Chen} \& {Han}}{2008}]{ch08}
{Chen} X.,  {Han} Z.,  2008, \mn@doi [\mnras]
  {10.1111/j.1365-2966.2008.13334.x}, \href
  {http://adsabs.harvard.edu/abs/2008MNRAS.387.1416C} {387, 1416}

\bibitem[\protect\citeauthoryear{{Chen}, {Woods}, {Yungelson}, {Gilfanov}  \&
  {Han}}{{Chen} et~al.}{2014}]{cwyg+14}
{Chen} H.-L.,  {Woods} T.~E.,  {Yungelson} L.~R.,  {Gilfanov} M.,   {Han} Z.,
  2014, \mn@doi [\mnras] {10.1093/mnras/stu1884}, \href
  {http://adsabs.harvard.edu/abs/2014MNRAS.445.1912C} {445, 1912}

\bibitem[\protect\citeauthoryear{{Chen}, {Woods}, {Yungelson}, {Gilfanov}  \&
  {Han}}{{Chen} et~al.}{2015}]{cwyg+15}
{Chen} H.-L.,  {Woods} T.~E.,  {Yungelson} L.~R.,  {Gilfanov} M.,   {Han} Z.,
  2015, \mn@doi [\mnras] {10.1093/mnras/stv1865}, \href
  {http://adsabs.harvard.edu/abs/2015MNRAS.453.3024C} {453, 3024}

\bibitem[\protect\citeauthoryear{{Ciardullo}, {Ford}, {Neill}, {Jacoby}  \&
  {Shafter}}{{Ciardullo} et~al.}{1987}]{cfnj+87}
{Ciardullo} R.,  {Ford} H.~C.,  {Neill} J.~D.,  {Jacoby} G.~H.,   {Shafter}
  A.~W.,  1987, \mn@doi [\apj] {10.1086/165388}, \href
  {http://adsabs.harvard.edu/abs/1987ApJ...318..520C} {318, 520}

\bibitem[\protect\citeauthoryear{{Ciardullo}, {Tamblyn}, {Jacoby}, {Ford}  \&
  {Williams}}{{Ciardullo} et~al.}{1990}]{ctjf+90}
{Ciardullo} R.,  {Tamblyn} P.,  {Jacoby} G.~H.,  {Ford} H.~C.,   {Williams}
  R.~E.,  1990, \mn@doi [\aj] {10.1086/115397}, \href
  {http://adsabs.harvard.edu/abs/1990AJ.....99.1079C} {99, 1079}

\bibitem[\protect\citeauthoryear{{Clark}}{{Clark}}{1975}]{clar75}
{Clark} G.~W.,  1975, \mn@doi [\apjl] {10.1086/181869}, \href
  {http://adsabs.harvard.edu/abs/1975ApJ...199L.143C} {199, L143}

\bibitem[\protect\citeauthoryear{{Coelho}, {Shafter}  \& {Misselt}}{{Coelho}
  et~al.}{2008}]{csm08}
{Coelho} E.~A.,  {Shafter} A.~W.,   {Misselt} K.~A.,  2008, \mn@doi [\apj]
  {10.1086/591517}, \href {http://adsabs.harvard.edu/abs/2008ApJ...686.1261C}
  {686, 1261}

\bibitem[\protect\citeauthoryear{{Curtin}, {Shafter}, {Pritchet}, {Neill},
  {Kundu}  \& {Maccarone}}{{Curtin} et~al.}{2015}]{cspn+15}
{Curtin} C.,  {Shafter} A.~W.,  {Pritchet} C.~J.,  {Neill} J.~D.,  {Kundu} A.,
   {Maccarone} T.~J.,  2015, \mn@doi [\apj] {10.1088/0004-637X/811/1/34}, \href
  {http://adsabs.harvard.edu/abs/2015ApJ...811...34C} {811, 34}

\bibitem[\protect\citeauthoryear{{Darnley} et~al.,}{{Darnley}
  et~al.}{2006}]{dbkn+06}
{Darnley} M.~J.,  et~al., 2006, \mn@doi [\mnras]
  {10.1111/j.1365-2966.2006.10297.x}, \href
  {http://adsabs.harvard.edu/abs/2006MNRAS.369..257D} {369, 257}

\bibitem[\protect\citeauthoryear{{Darnley} et~al.,}{{Darnley}
  et~al.}{2014}]{dbhh+14}
{Darnley} M.~J.,  et~al., 2014, in {Woudt} P.~A.,  {Ribeiro} V.~A.~R.~M.,  eds,
   Astronomical Society of the Pacific Conference Series Vol. 490, Stell Novae:
  Past and Future Decades. p.~49 (\mn@eprint {arXiv} {1303.2711})

\bibitem[\protect\citeauthoryear{{Davis}, {Kolb}  \& {Willems}}{{Davis}
  et~al.}{2010}]{dkw10}
{Davis} P.~J.,  {Kolb} U.,   {Willems} B.,  2010, \mn@doi [\mnras]
  {10.1111/j.1365-2966.2009.16138.x}, \href
  {http://adsabs.harvard.edu/abs/2010MNRAS.403..179D} {403, 179}

\bibitem[\protect\citeauthoryear{{Davis}, {Kolb}  \& {Knigge}}{{Davis}
  et~al.}{2012}]{dkk12}
{Davis} P.~J.,  {Kolb} U.,   {Knigge} C.,  2012, \mn@doi [\mnras]
  {10.1111/j.1365-2966.2011.19690.x}, \href
  {http://adsabs.harvard.edu/abs/2012MNRAS.419..287D} {419, 287}

\bibitem[\protect\citeauthoryear{{de Kool}}{{de Kool}}{1990}]{deko90}
{de Kool} M.,  1990, \mn@doi [\apj] {10.1086/168974}, \href
  {http://adsabs.harvard.edu/abs/1990ApJ...358..189D} {358, 189}

\bibitem[\protect\citeauthoryear{{Della Valle}}{{Della Valle}}{2002}]{dell02}
{Della Valle} M.,  2002, in {Hernanz} M.,  {Jos{\'e}} J.,  eds,  American
  Institute of Physics Conference Series Vol. 637, Classical Nova Explosions.
  pp 443--456 (\mn@eprint {} {astro-ph/0210276}), \mn@doi{10.1063/1.1518244}

\bibitem[\protect\citeauthoryear{{Della Valle} \& {Livio}}{{Della Valle} \&
  {Livio}}{1998}]{dl98}
{Della Valle} M.,  {Livio} M.,  1998, \mn@doi [\apj] {10.1086/306275}, \href
  {http://adsabs.harvard.edu/abs/1998ApJ...506..818D} {506, 818}

\bibitem[\protect\citeauthoryear{{Della Valle}, {Bianchini}, {Livio}  \&
  {Orio}}{{Della Valle} et~al.}{1992}]{dblo92}
{Della Valle} M.,  {Bianchini} A.,  {Livio} M.,   {Orio} M.,  1992, \aap, \href
  {http://adsabs.harvard.edu/abs/1992A%26A...266..232D} {266, 232}

\bibitem[\protect\citeauthoryear{{Della Valle}, {Rosino}, {Bianchini}  \&
  {Livio}}{{Della Valle} et~al.}{1994}]{drbl94}
{Della Valle} M.,  {Rosino} L.,  {Bianchini} A.,   {Livio} M.,  1994, \aap,
  \href {http://adsabs.harvard.edu/abs/1994A%26A...287..403D} {287, 403}

\bibitem[\protect\citeauthoryear{{Dieball}, {Long}, {Knigge}, {Thomson}  \&
  {Zurek}}{{Dieball} et~al.}{2010}]{dlkt+10}
{Dieball} A.,  {Long} K.~S.,  {Knigge} C.,  {Thomson} G.~S.,   {Zurek} D.~R.,
  2010, \mn@doi [\apj] {10.1088/0004-637X/710/1/332}, \href
  {http://adsabs.harvard.edu/abs/2010ApJ...710..332D} {710, 332}

\bibitem[\protect\citeauthoryear{{Duerbeck}}{{Duerbeck}}{1990}]{duer90}
{Duerbeck} H.~W.,  1990, in {Cassatella} A.,  {Viotti} R.,  eds,  Lecture Notes
  in Physics, Berlin Springer Verlag Vol. 369, IAU Colloq. 122: Physics of
  Classical Novae. p.~34, \mn@doi{10.1007/3-540-53500-4_90}

\bibitem[\protect\citeauthoryear{{Ferrarese}, {C{\^o}t{\'e}}  \&
  {Jord{\'a}n}}{{Ferrarese} et~al.}{2003}]{fcj03}
{Ferrarese} L.,  {C{\^o}t{\'e}} P.,   {Jord{\'a}n} A.,  2003, \mn@doi [\apj]
  {10.1086/379349}, \href {http://adsabs.harvard.edu/abs/2003ApJ...599.1302F}
  {599, 1302}

\bibitem[\protect\citeauthoryear{{Franck}, {Shafter}, {Hornoch}  \&
  {Misselt}}{{Franck} et~al.}{2012}]{fshm12}
{Franck} J.~R.,  {Shafter} A.~W.,  {Hornoch} K.,   {Misselt} K.~A.,  2012,
  \mn@doi [\apj] {10.1088/0004-637X/760/1/13}, \href
  {http://adsabs.harvard.edu/abs/2012ApJ...760...13F} {760, 13}

\bibitem[\protect\citeauthoryear{{Gehrz}, {Truran}, {Williams}  \&
  {Starrfield}}{{Gehrz} et~al.}{1998}]{gtws98}
{Gehrz} R.~D.,  {Truran} J.~W.,  {Williams} R.~E.,   {Starrfield} S.,  1998,
  \mn@doi [\pasp] {10.1086/316107}, \href
  {http://adsabs.harvard.edu/abs/1998PASP..110....3G} {110, 3}

\bibitem[\protect\citeauthoryear{{Giannone} \& {Weigert}}{{Giannone} \&
  {Weigert}}{1967}]{gw67}
{Giannone} P.,  {Weigert} A.,  1967, \zap, \href
  {http://adsabs.harvard.edu/abs/1967ZA.....67...41G} {67, 41}

\bibitem[\protect\citeauthoryear{{Grindlay}, {Heinke}, {Edmonds}  \&
  {Murray}}{{Grindlay} et~al.}{2001}]{ghem01}
{Grindlay} J.~E.,  {Heinke} C.,  {Edmonds} P.~D.,   {Murray} S.~S.,  2001,
  \mn@doi [Science] {10.1126/science.1061135}, \href
  {http://adsabs.harvard.edu/abs/2001Sci...292.2290G} {292, 2290}

\bibitem[\protect\citeauthoryear{{Hachisu} \& {Kato}}{{Hachisu} \&
  {Kato}}{2001}]{hk01}
{Hachisu} I.,  {Kato} M.,  2001, \mn@doi [\apj] {10.1086/321601}, \href
  {http://adsabs.harvard.edu/abs/2001ApJ...558..323H} {558, 323}

\bibitem[\protect\citeauthoryear{{Hachisu}, {Kato}  \& {Nomoto}}{{Hachisu}
  et~al.}{1996}]{hkn96}
{Hachisu} I.,  {Kato} M.,   {Nomoto} K.,  1996, \mn@doi [\apjl]
  {10.1086/310303}, \href {http://adsabs.harvard.edu/abs/1996ApJ...470L..97H}
  {470, L97}

\bibitem[\protect\citeauthoryear{{Hachisu}, {Kato}  \& {Nomoto}}{{Hachisu}
  et~al.}{1999}]{hkn99}
{Hachisu} I.,  {Kato} M.,   {Nomoto} K.,  1999, \mn@doi [\apj]
  {10.1086/307608}, \href {http://adsabs.harvard.edu/abs/1999ApJ...522..487H}
  {522, 487}

\bibitem[\protect\citeauthoryear{{Han}, {Podsiadlowski}, {Maxted}, {Marsh}  \&
  {Ivanova}}{{Han} et~al.}{2002}]{hpmm+02}
{Han} Z.,  {Podsiadlowski} P.,  {Maxted} P.~F.~L.,  {Marsh} T.~R.,   {Ivanova}
  N.,  2002, \mn@doi [\mnras] {10.1046/j.1365-8711.2002.05752.x}, \href
  {http://adsabs.harvard.edu/abs/2002MNRAS.336..449H} {336, 449}

\bibitem[\protect\citeauthoryear{{Heinke}, {Grindlay}, {Edmonds}, {Cohn},
  {Lugger}, {Camilo}, {Bogdanov}  \& {Freire}}{{Heinke} et~al.}{2005}]{hgec+05}
{Heinke} C.~O.,  {Grindlay} J.~E.,  {Edmonds} P.~D.,  {Cohn} H.~N.,  {Lugger}
  P.~M.,  {Camilo} F.,  {Bogdanov} S.,   {Freire} P.~C.,  2005, \mn@doi [\apj]
  {10.1086/429899}, \href {http://adsabs.harvard.edu/abs/2005ApJ...625..796H}
  {625, 796}

\bibitem[\protect\citeauthoryear{{Henze} et~al.,}{{Henze}
  et~al.}{2013}]{hphd+13}
{Henze} M.,  et~al., 2013, \mn@doi [\aap] {10.1051/0004-6361/201220196}, \href
  {http://adsabs.harvard.edu/abs/2013A%26A...549A.120H} {549, A120}

\bibitem[\protect\citeauthoryear{{Henze}, {Darnley}, {Kabashima}, {Nishiyama},
  {Itagaki}  \& {Gao}}{{Henze} et~al.}{2015}]{hdkn+15}
{Henze} M.,  {Darnley} M.~J.,  {Kabashima} F.,  {Nishiyama} K.,  {Itagaki} K.,
   {Gao} X.,  2015, \mn@doi [\aap] {10.1051/0004-6361/201527168}, \href
  {http://adsabs.harvard.edu/abs/2015A%26A...582L...8H} {582, L8}

\bibitem[\protect\citeauthoryear{{Hernanz}, {Jose}, {Coc}  \&
  {Isern}}{{Hernanz} et~al.}{1996}]{hjci96}
{Hernanz} M.,  {Jose} J.,  {Coc} A.,   {Isern} J.,  1996, \mn@doi [\apjl]
  {10.1086/310122}, \href {http://adsabs.harvard.edu/abs/1996ApJ...465L..27H}
  {465, L27}

\bibitem[\protect\citeauthoryear{{Hillman}, {Prialnik}, {Kovetz}  \&
  {Shara}}{{Hillman} et~al.}{2015}]{hpks15}
{Hillman} Y.,  {Prialnik} D.,  {Kovetz} A.,   {Shara} M.~M.,  2015, preprint,
  \href {http://adsabs.harvard.edu/abs/2015arXiv150803141H} {} (\mn@eprint
  {arXiv} {1508.03141})

\bibitem[\protect\citeauthoryear{{Hjellming} \& {Webbink}}{{Hjellming} \&
  {Webbink}}{1987}]{hw87}
{Hjellming} M.~S.,  {Webbink} R.~F.,  1987, \mn@doi [\apj] {10.1086/165412},
  \href {http://adsabs.harvard.edu/abs/1987ApJ...318..794H} {318, 794}

\bibitem[\protect\citeauthoryear{{Hurley}, {Tout}  \& {Pols}}{{Hurley}
  et~al.}{2002}]{htp02}
{Hurley} J.~R.,  {Tout} C.~A.,   {Pols} O.~R.,  2002, \mn@doi [\mnras]
  {10.1046/j.1365-8711.2002.05038.x}, \href
  {http://adsabs.harvard.edu/abs/2002MNRAS.329..897H} {329, 897}

\bibitem[\protect\citeauthoryear{{Idan}, {Shaviv}  \& {Shaviv}}{{Idan}
  et~al.}{2013}]{iss13}
{Idan} I.,  {Shaviv} N.~J.,   {Shaviv} G.,  2013, \mn@doi [\mnras]
  {10.1093/mnras/stt908}, \href
  {http://adsabs.harvard.edu/abs/2013MNRAS.433.2884I} {433, 2884}

\bibitem[\protect\citeauthoryear{{Ivanova} et~al.,}{{Ivanova}
  et~al.}{2013}]{ijcd+13}
{Ivanova} N.,  et~al., 2013, \mn@doi [\aapr] {10.1007/s00159-013-0059-2}, \href
  {http://adsabs.harvard.edu/abs/2013A%26ARv..21...59I} {21, 59}

\bibitem[\protect\citeauthoryear{{Izzo} et~al.,}{{Izzo} et~al.}{2015}]{idmm+15}
{Izzo} L.,  et~al., 2015, \mn@doi [\apjl] {10.1088/2041-8205/808/1/L14}, \href
  {http://adsabs.harvard.edu/abs/2015ApJ...808L..14I} {808, L14}

\bibitem[\protect\citeauthoryear{{Johnson}}{{Johnson}}{1966}]{john66}
{Johnson} H.~L.,  1966, \mn@doi [\araa] {10.1146/annurev.aa.04.090166.001205},
  \href {http://adsabs.harvard.edu/abs/1966ARA%26A...4..193J} {4, 193}

\bibitem[\protect\citeauthoryear{{Jos{\'e}} \& {Hernanz}}{{Jos{\'e}} \&
  {Hernanz}}{1998}]{jh98}
{Jos{\'e}} J.,  {Hernanz} M.,  1998, \mn@doi [\apj] {10.1086/305244}, \href
  {http://adsabs.harvard.edu/abs/1998ApJ...494..680J} {494, 680}

\bibitem[\protect\citeauthoryear{{Kato} \& {Hachisu}}{{Kato} \&
  {Hachisu}}{2004}]{kh04}
{Kato} M.,  {Hachisu} I.,  2004, \mn@doi [\apjl] {10.1086/425249}, \href
  {http://adsabs.harvard.edu/abs/2004ApJ...613L.129K} {613, L129}

\bibitem[\protect\citeauthoryear{{Kouwenhoven}, {Brown}, {Goodwin}, {Portegies
  Zwart}  \& {Kaper}}{{Kouwenhoven} et~al.}{2009}]{kbgp+09}
{Kouwenhoven} M.~B.~N.,  {Brown} A.~G.~A.,  {Goodwin} S.~P.,  {Portegies Zwart}
  S.~F.,   {Kaper} L.,  2009, \mn@doi [\aap] {10.1051/0004-6361:200810234},
  \href {http://adsabs.harvard.edu/abs/2009A%26A...493..979K} {493, 979}

\bibitem[\protect\citeauthoryear{{Kraft}}{{Kraft}}{1964}]{kraf64}
{Kraft} R.~P.,  1964, \mn@doi [\apj] {10.1086/147776}, \href
  {http://adsabs.harvard.edu/abs/1964ApJ...139..457K} {139, 457}

\bibitem[\protect\citeauthoryear{{Kraicheva}, {Popova}, {Tutukov}  \&
  {Yungelson}}{{Kraicheva} et~al.}{1979}]{kpty79}
{Kraicheva} Z.~T.,  {Popova} E.~I.,  {Tutukov} A.~V.,   {Yungelson} L.~R.,
  1979, \sovast, \href {http://adsabs.harvard.edu/abs/1979SvA....23..290K} {23,
  290}

\bibitem[\protect\citeauthoryear{{Kraus} \& {Hillenbrand}}{{Kraus} \&
  {Hillenbrand}}{2009}]{kh09}
{Kraus} A.~L.,  {Hillenbrand} L.~A.,  2009, \mn@doi [\apj]
  {10.1088/0004-637X/703/2/1511}, \href
  {http://adsabs.harvard.edu/abs/2009ApJ...703.1511K} {703, 1511}

\bibitem[\protect\citeauthoryear{{Kroupa}}{{Kroupa}}{2001}]{krou01}
{Kroupa} P.,  2001, \mn@doi [\mnras] {10.1046/j.1365-8711.2001.04022.x}, \href
  {http://adsabs.harvard.edu/abs/2001MNRAS.322..231K} {322, 231}

\bibitem[\protect\citeauthoryear{{Kudryashov}, {Chuga{\u i}}  \&
  {Tutukov}}{{Kudryashov} et~al.}{2000}]{kct00}
{Kudryashov} A.~D.,  {Chuga{\u i}} N.~N.,   {Tutukov} A.~V.,  2000, \mn@doi
  [Astronomy Reports] {10.1134/1.163838}, \href
  {http://adsabs.harvard.edu/abs/2000ARep...44..170K} {44, 170}

\bibitem[\protect\citeauthoryear{{Livio}, {Govarie}  \& {Ritter}}{{Livio}
  et~al.}{1991}]{lgr91}
{Livio} M.,  {Govarie} A.,   {Ritter} H.,  1991, \aap, \href
  {http://adsabs.harvard.edu/abs/1991A%26A...246...84L} {246, 84}

\bibitem[\protect\citeauthoryear{{Loveridge}, {van der Sluys}  \&
  {Kalogera}}{{Loveridge} et~al.}{2011}]{lvk11}
{Loveridge} A.~J.,  {van der Sluys} M.~V.,   {Kalogera} V.,  2011, \mn@doi
  [\apj] {10.1088/0004-637X/743/1/49}, \href
  {http://adsabs.harvard.edu/abs/2011ApJ...743...49L} {743, 49}

\bibitem[\protect\citeauthoryear{{Matteucci}, {Renda}, {Pipino}  \& {Della
  Valle}}{{Matteucci} et~al.}{2003}]{mrpd03}
{Matteucci} F.,  {Renda} A.,  {Pipino} A.,   {Della Valle} M.,  2003, \mn@doi
  [\aap] {10.1051/0004-6361:20030520}, \href
  {http://adsabs.harvard.edu/abs/2003A%26A...405...23M} {405, 23}

\bibitem[\protect\citeauthoryear{{Meng}, {Chen}  \& {Han}}{{Meng}
  et~al.}{2008}]{mch08}
{Meng} X.,  {Chen} X.,   {Han} Z.,  2008, \mn@doi [\aap]
  {10.1051/0004-6361:20078841}, \href
  {http://adsabs.harvard.edu/abs/2008A%26A...487..625M} {487, 625}

\bibitem[\protect\citeauthoryear{{Mestel}}{{Mestel}}{1952a}]{mest52a}
{Mestel} L.,  1952a, \mnras, \href
  {http://adsabs.harvard.edu/abs/1952MNRAS.112..583M} {112, 583}

\bibitem[\protect\citeauthoryear{{Mestel}}{{Mestel}}{1952b}]{mest52b}
{Mestel} L.,  1952b, \mnras, \href
  {http://adsabs.harvard.edu/abs/1952MNRAS.112..598M} {112, 598}

\bibitem[\protect\citeauthoryear{{Munari}}{{Munari}}{2014}]{muna14}
{Munari} U.,  2014, in {Woudt} P.~A.,  {Ribeiro} V.~A.~R.~M.,  eds,
  Astronomical Society of the Pacific Conference Series Vol. 490, Stell Novae:
  Past and Future Decades. p.~183

\bibitem[\protect\citeauthoryear{{Neill} \& {Shara}}{{Neill} \&
  {Shara}}{2005}]{ns05}
{Neill} J.~D.,  {Shara} M.~M.,  2005, \mn@doi [\aj] {10.1086/428482}, \href
  {http://adsabs.harvard.edu/abs/2005AJ....129.1873N} {129, 1873}

\bibitem[\protect\citeauthoryear{{Nelemans}, {Siess}, {Repetto}, {Toonen}  \&
  {Phinney}}{{Nelemans} et~al.}{2016}]{nsrt+16}
{Nelemans} G.,  {Siess} L.,  {Repetto} S.,  {Toonen} S.,   {Phinney} E.~S.,
  2016, \mn@doi [\apj] {10.3847/0004-637X/817/1/69}, \href
  {http://adsabs.harvard.edu/abs/2016ApJ...817...69N} {817, 69}

\bibitem[\protect\citeauthoryear{{Nelson}, {MacCannell}  \& {Dubeau}}{{Nelson}
  et~al.}{2004}]{nmd04}
{Nelson} L.~A.,  {MacCannell} K.~A.,   {Dubeau} E.,  2004, \mn@doi [\apj]
  {10.1086/381156}, \href {http://adsabs.harvard.edu/abs/2004ApJ...602..938N}
  {602, 938}

\bibitem[\protect\citeauthoryear{{Olsen}, {Blum}, {Stephens}, {Davidge},
  {Massey}, {Strom}  \& {Rigaut}}{{Olsen} et~al.}{2006}]{obsd+06}
{Olsen} K.~A.~G.,  {Blum} R.~D.,  {Stephens} A.~W.,  {Davidge} T.~J.,  {Massey}
  P.,  {Strom} S.~E.,   {Rigaut} F.,  2006, \mn@doi [\aj] {10.1086/504900},
  \href {http://adsabs.harvard.edu/abs/2006AJ....132..271O} {132, 271}

\bibitem[\protect\citeauthoryear{{Paczynski} \& {Zytkow}}{{Paczynski} \&
  {Zytkow}}{1978}]{pz78}
{Paczynski} B.,  {Zytkow} A.~N.,  1978, \mn@doi [\apj] {10.1086/156176}, \href
  {http://adsabs.harvard.edu/abs/1978ApJ...222..604P} {222, 604}

\bibitem[\protect\citeauthoryear{{Passy}, {Herwig}  \& {Paxton}}{{Passy}
  et~al.}{2012}]{php12}
{Passy} J.-C.,  {Herwig} F.,   {Paxton} B.,  2012, \mn@doi [\apj]
  {10.1088/0004-637X/760/1/90}, \href
  {http://adsabs.harvard.edu/abs/2012ApJ...760...90P} {760, 90}

\bibitem[\protect\citeauthoryear{{Patterson}}{{Patterson}}{1984}]{patt84}
{Patterson} J.,  1984, \mn@doi [\apjs] {10.1086/190940}, \href
  {http://adsabs.harvard.edu/abs/1984ApJS...54..443P} {54, 443}

\bibitem[\protect\citeauthoryear{{Pavlovskii} \& {Ivanova}}{{Pavlovskii} \&
  {Ivanova}}{2015}]{pi15}
{Pavlovskii} K.,  {Ivanova} N.,  2015, \mn@doi [\mnras] {10.1093/mnras/stv619},
  \href {http://adsabs.harvard.edu/abs/2015MNRAS.449.4415P} {449, 4415}

\bibitem[\protect\citeauthoryear{{Paxton}, {Bildsten}, {Dotter}, {Herwig},
  {Lesaffre}  \& {Timmes}}{{Paxton} et~al.}{2011}]{pbdh+11}
{Paxton} B.,  {Bildsten} L.,  {Dotter} A.,  {Herwig} F.,  {Lesaffre} P.,
  {Timmes} F.,  2011, \mn@doi [\apjs] {10.1088/0067-0049/192/1/3}, \href
  {http://adsabs.harvard.edu/abs/2011ApJS..192....3P} {192, 3}

\bibitem[\protect\citeauthoryear{{Paxton} et~al.,}{{Paxton}
  et~al.}{2013}]{pcab+13}
{Paxton} B.,  et~al., 2013, \mn@doi [\apjs] {10.1088/0067-0049/208/1/4}, \href
  {http://adsabs.harvard.edu/abs/2013ApJS..208....4P} {208, 4}

\bibitem[\protect\citeauthoryear{{Piersanti}, {Cassisi}, {Iben}  \&
  {Tornamb{\'e}}}{{Piersanti} et~al.}{2000}]{pcit00}
{Piersanti} L.,  {Cassisi} S.,  {Iben} Jr. I.,   {Tornamb{\'e}} A.,  2000,
  \mn@doi [\apj] {10.1086/308885}, \href
  {http://adsabs.harvard.edu/abs/2000ApJ...535..932P} {535, 932}

\bibitem[\protect\citeauthoryear{{Piersanti}, {Tornamb{\'e}}  \&
  {Yungelson}}{{Piersanti} et~al.}{2014}]{pty14}
{Piersanti} L.,  {Tornamb{\'e}} A.,   {Yungelson} L.~R.,  2014, \mn@doi
  [\mnras] {10.1093/mnras/stu1885}, \href
  {http://adsabs.harvard.edu/abs/2014MNRAS.445.3239P} {445, 3239}

\bibitem[\protect\citeauthoryear{{Politano}}{{Politano}}{1996}]{poli96}
{Politano} M.,  1996, \mn@doi [\apj] {10.1086/177423}, \href
  {http://adsabs.harvard.edu/abs/1996ApJ...465..338P} {465, 338}

\bibitem[\protect\citeauthoryear{{Politano}, {Livio}, {Truran}  \&
  {Webbink}}{{Politano} et~al.}{1990}]{pltw90}
{Politano} M.,  {Livio} M.,  {Truran} J.~W.,   {Webbink} R.~F.,  1990, in
  {Cassatella} A.,  {Viotti} R.,  eds,  Lecture Notes in Physics, Berlin
  Springer Verlag Vol. 369, IAU Colloq. 122: Physics of Classical Novae.
  p.~386, \mn@doi{10.1007/3-540-53500-4_150}

\bibitem[\protect\citeauthoryear{{Pooley} et~al.,}{{Pooley}
  et~al.}{2002}]{plhv+02}
{Pooley} D.,  et~al., 2002, \mn@doi [\apj] {10.1086/339210}, \href
  {http://adsabs.harvard.edu/abs/2002ApJ...569..405P} {569, 405}

\bibitem[\protect\citeauthoryear{{Prialnik} \& {Kovetz}}{{Prialnik} \&
  {Kovetz}}{1995}]{pk95}
{Prialnik} D.,  {Kovetz} A.,  1995, \mn@doi [\apj] {10.1086/175741}, \href
  {http://adsabs.harvard.edu/abs/1995ApJ...445..789P} {445, 789}

\bibitem[\protect\citeauthoryear{{Prialnik}, {Shara}  \& {Shaviv}}{{Prialnik}
  et~al.}{1978}]{pss78}
{Prialnik} D.,  {Shara} M.~M.,   {Shaviv} G.,  1978, \aap, \href
  {http://adsabs.harvard.edu/abs/1978A%26A....62..339P} {62, 339}

\bibitem[\protect\citeauthoryear{{Prialnik}, {Shara}  \& {Shaviv}}{{Prialnik}
  et~al.}{1979}]{pss79}
{Prialnik} D.,  {Shara} M.~M.,   {Shaviv} G.,  1979, \aap, \href
  {http://adsabs.harvard.edu/abs/1979A%26A....72..192P} {72, 192}

\bibitem[\protect\citeauthoryear{{Ricker} \& {Taam}}{{Ricker} \&
  {Taam}}{2012}]{rt12}
{Ricker} P.~M.,  {Taam} R.~E.,  2012, \mn@doi [\apj]
  {10.1088/0004-637X/746/1/74}, \href
  {http://adsabs.harvard.edu/abs/2012ApJ...746...74R} {746, 74}

\bibitem[\protect\citeauthoryear{{Robertson}, {Yoshida}, {Springel}  \&
  {Hernquist}}{{Robertson} et~al.}{2004}]{rysh04}
{Robertson} B.,  {Yoshida} N.,  {Springel} V.,   {Hernquist} L.,  2004, \mn@doi
  [\apj] {10.1086/382871}, \href
  {http://adsabs.harvard.edu/abs/2004ApJ...606...32R} {606, 32}

\bibitem[\protect\citeauthoryear{{Rosino}}{{Rosino}}{1964}]{rosi64}
{Rosino} L.,  1964, Annales d'Astrophysique, \href
  {http://adsabs.harvard.edu/abs/1964AnAp...27..498R} {27, 498}

\bibitem[\protect\citeauthoryear{{Rosino}}{{Rosino}}{1973}]{rosi73}
{Rosino} L.,  1973, \aaps, \href
  {http://adsabs.harvard.edu/abs/1973A%26AS....9..347R} {9, 347}

\bibitem[\protect\citeauthoryear{{Rosino}, {Capaccioli}, {D'Onofrio}  \& {Della
  Valle}}{{Rosino} et~al.}{1989}]{rcdd89}
{Rosino} L.,  {Capaccioli} M.,  {D'Onofrio} M.,   {Della Valle} M.,  1989,
  \mn@doi [\aj] {10.1086/114959}, \href
  {http://adsabs.harvard.edu/abs/1989AJ.....97...83R} {97, 83}

\bibitem[\protect\citeauthoryear{{Saglia} et~al.,}{{Saglia}
  et~al.}{2010}]{sfbm+10}
{Saglia} R.~P.,  et~al., 2010, \mn@doi [\aap] {10.1051/0004-6361/200912805},
  \href {http://adsabs.harvard.edu/abs/2010A%26A...509A..61S} {509, A61}

\bibitem[\protect\citeauthoryear{{Sana} et~al.,}{{Sana} et~al.}{2012}]{sddl+12}
{Sana} H.,  et~al., 2012, \mn@doi [Science] {10.1126/science.1223344}, \href
  {http://adsabs.harvard.edu/abs/2012Sci...337..444S} {337, 444}

\bibitem[\protect\citeauthoryear{{Schreiber}, {Zorotovic}  \&
  {Wijnen}}{{Schreiber} et~al.}{2016}]{szw16}
{Schreiber} M.~R.,  {Zorotovic} M.,   {Wijnen} T.~P.~G.,  2016, \mn@doi
  [\mnras] {10.1093/mnrasl/slv144}, \href
  {http://adsabs.harvard.edu/abs/2016MNRAS.455L..16S} {455, L16}

\bibitem[\protect\citeauthoryear{{Schwartzman}, {Kovetz}  \&
  {Prialnik}}{{Schwartzman} et~al.}{1994}]{skp94}
{Schwartzman} E.,  {Kovetz} A.,   {Prialnik} D.,  1994, \mnras, \href
  {http://adsabs.harvard.edu/abs/1994MNRAS.269..323S} {269, 323}

\bibitem[\protect\citeauthoryear{{Shafter} \& {Irby}}{{Shafter} \&
  {Irby}}{2001}]{si01}
{Shafter} A.~W.,  {Irby} B.~K.,  2001, \mn@doi [\apj] {10.1086/324044}, \href
  {http://adsabs.harvard.edu/abs/2001ApJ...563..749S} {563, 749}

\bibitem[\protect\citeauthoryear{{Shafter} \& {Quimby}}{{Shafter} \&
  {Quimby}}{2007}]{sq07}
{Shafter} A.~W.,  {Quimby} R.~M.,  2007, \mn@doi [\apjl] {10.1086/525254},
  \href {http://adsabs.harvard.edu/abs/2007ApJ...671L.121S} {671, L121}

\bibitem[\protect\citeauthoryear{{Shafter}, {Ciardullo}  \&
  {Pritchet}}{{Shafter} et~al.}{2000}]{scp00}
{Shafter} A.~W.,  {Ciardullo} R.,   {Pritchet} C.~J.,  2000, \mn@doi [\apj]
  {10.1086/308349}, \href {http://adsabs.harvard.edu/abs/2000ApJ...530..193S}
  {530, 193}

\bibitem[\protect\citeauthoryear{{Shafter}, {Rau}, {Quimby}, {Kasliwal},
  {Bode}, {Darnley}  \& {Misselt}}{{Shafter} et~al.}{2009}]{srqk+09}
{Shafter} A.~W.,  {Rau} A.,  {Quimby} R.~M.,  {Kasliwal} M.~M.,  {Bode} M.~F.,
  {Darnley} M.~J.,   {Misselt} K.~A.,  2009, \mn@doi [\apj]
  {10.1088/0004-637X/690/2/1148}, \href
  {http://adsabs.harvard.edu/abs/2009ApJ...690.1148S} {690, 1148}

\bibitem[\protect\citeauthoryear{{Shafter} et~al.,}{{Shafter}
  et~al.}{2015}]{shrs+15}
{Shafter} A.~W.,  et~al., 2015, \mn@doi [\apjs] {10.1088/0067-0049/216/2/34},
  \href {http://adsabs.harvard.edu/abs/2015ApJS..216...34S} {216, 34}

\bibitem[\protect\citeauthoryear{{Shara}, {Zurek}, {Baltz}, {Lauer}  \&
  {Silk}}{{Shara} et~al.}{2004}]{szbl+04}
{Shara} M.~M.,  {Zurek} D.~R.,  {Baltz} E.~A.,  {Lauer} T.~R.,   {Silk} J.,
  2004, \mn@doi [\apjl] {10.1086/420882}, \href
  {http://adsabs.harvard.edu/abs/2004ApJ...605L.117S} {605, L117}

\bibitem[\protect\citeauthoryear{{Shara}, {Yaron}, {Prialnik}, {Kovetz}  \&
  {Zurek}}{{Shara} et~al.}{2010}]{sypk+10}
{Shara} M.~M.,  {Yaron} O.,  {Prialnik} D.,  {Kovetz} A.,   {Zurek} D.,  2010,
  \mn@doi [\apj] {10.1088/0004-637X/725/1/831}, \href
  {http://adsabs.harvard.edu/abs/2010ApJ...725..831S} {725, 831}

\bibitem[\protect\citeauthoryear{{Shara}, {Doyle}, {Lauer}, {Zurek}, {Neill},
  {Madrid}, {Welch}  \& {Baltz}}{{Shara} et~al.}{2016}]{sdlz+16}
{Shara} M.~M.,  {Doyle} T.,  {Lauer} T.~R.,  {Zurek} D.,  {Neill} J.~D.,
  {Madrid} J.~P.,  {Welch} D.~L.,   {Baltz} E.~A.,  2016, preprint, \href
  {http://adsabs.harvard.edu/abs/2016arXiv160200758S} {} (\mn@eprint {arXiv}
  {1602.00758})

\bibitem[\protect\citeauthoryear{{Sion}, {Acierno}  \& {Tomczyk}}{{Sion}
  et~al.}{1979}]{sat79}
{Sion} E.~M.,  {Acierno} M.~J.,   {Tomczyk} S.,  1979, \mn@doi [\apj]
  {10.1086/157143}, \href {http://adsabs.harvard.edu/abs/1979ApJ...230..832S}
  {230, 832}

\bibitem[\protect\citeauthoryear{{Soraisam} \& {Gilfanov}}{{Soraisam} \&
  {Gilfanov}}{2015}]{sg14}
{Soraisam} M.~D.,  {Gilfanov} M.,  2015, \mn@doi [\aap]
  {10.1051/0004-6361/201424118}, \href
  {http://adsabs.harvard.edu/abs/2015A%26A...583A.140S} {583, A140}

\bibitem[\protect\citeauthoryear{{Soraisam}, {Gilfanov}, {Wolf}  \&
  {Bildsten}}{{Soraisam} et~al.}{2016}]{sgwb15}
{Soraisam} M.~D.,  {Gilfanov} M.,  {Wolf} W.~M.,   {Bildsten} L.,  2016,
  \mn@doi [\mnras] {10.1093/mnras/stv2359}, \href
  {http://adsabs.harvard.edu/abs/2016MNRAS.455..668S} {455, 668}

\bibitem[\protect\citeauthoryear{{Starrfield}, {Truran}, {Sparks}  \&
  {Kutter}}{{Starrfield} et~al.}{1972}]{stsk72}
{Starrfield} S.,  {Truran} J.~W.,  {Sparks} W.~M.,   {Kutter} G.~S.,  1972,
  \mn@doi [\apj] {10.1086/151619}, \href
  {http://adsabs.harvard.edu/abs/1972ApJ...176..169S} {176, 169}

\bibitem[\protect\citeauthoryear{{Starrfield}, {Truran}  \&
  {Sparks}}{{Starrfield} et~al.}{1978}]{sts78}
{Starrfield} S.,  {Truran} J.~W.,   {Sparks} W.~M.,  1978, \mn@doi [\apj]
  {10.1086/156598}, \href {http://adsabs.harvard.edu/abs/1978ApJ...226..186S}
  {226, 186}

\bibitem[\protect\citeauthoryear{{Starrfield}, {Sparks}  \&
  {Shaviv}}{{Starrfield} et~al.}{1988}]{sss88}
{Starrfield} S.,  {Sparks} W.~M.,   {Shaviv} G.,  1988, \mn@doi [\apjl]
  {10.1086/185105}, \href {http://adsabs.harvard.edu/abs/1988ApJ...325L..35S}
  {325, L35}

\bibitem[\protect\citeauthoryear{{Starrfield}, {Schwarz}, {Truran}  \&
  {Sparks}}{{Starrfield} et~al.}{2000}]{ssts00}
{Starrfield} S.,  {Schwarz} G.,  {Truran} J.~W.,   {Sparks} W.~M.,  2000, in
  {Holt} S.~S.,  {Zhang} W.~W.,  eds,  American Institute of Physics Conference
  Series Vol. 522, American Institute of Physics Conference Series. pp
  379--382, \mn@doi{10.1063/1.1291739}

\bibitem[\protect\citeauthoryear{{Tajitsu}, {Sadakane}, {Naito}, {Arai}  \&
  {Aoki}}{{Tajitsu} et~al.}{2015}]{tsna+15}
{Tajitsu} A.,  {Sadakane} K.,  {Naito} H.,  {Arai} A.,   {Aoki} W.,  2015,
  \mn@doi [\nat] {10.1038/nature14161}, \href
  {http://adsabs.harvard.edu/abs/2015Natur.518..381T} {518, 381}

\bibitem[\protect\citeauthoryear{{Tajitsu}, {Sadakane}, {Naito}, {Arai},
  {Kawakita}  \& {Aoki}}{{Tajitsu} et~al.}{2016}]{tsna+16}
{Tajitsu} A.,  {Sadakane} K.,  {Naito} H.,  {Arai} A.,  {Kawakita} H.,   {Aoki}
  W.,  2016, preprint, \href
  {http://adsabs.harvard.edu/abs/2016arXiv160105168T} {} (\mn@eprint {arXiv}
  {1601.05168})

\bibitem[\protect\citeauthoryear{{Tang} et~al.,}{{Tang} et~al.}{2014}]{tbwl+14}
{Tang} S.,  et~al., 2014, \mn@doi [\apj] {10.1088/0004-637X/786/1/61}, \href
  {http://adsabs.harvard.edu/abs/2014ApJ...786...61T} {786, 61}

\bibitem[\protect\citeauthoryear{{Townsley} \& {Bildsten}}{{Townsley} \&
  {Bildsten}}{2004}]{tb04}
{Townsley} D.~M.,  {Bildsten} L.,  2004, \mn@doi [\apj] {10.1086/379647}, \href
  {http://adsabs.harvard.edu/abs/2004ApJ...600..390T} {600, 390}

\bibitem[\protect\citeauthoryear{{Truran} \& {Livio}}{{Truran} \&
  {Livio}}{1986}]{tl86}
{Truran} J.~W.,  {Livio} M.,  1986, \mn@doi [\apj] {10.1086/164544}, \href
  {http://adsabs.harvard.edu/abs/1986ApJ...308..721T} {308, 721}

\bibitem[\protect\citeauthoryear{{Tutukov} \& {Ergma}}{{Tutukov} \&
  {Ergma}}{1979}]{te79}
{Tutukov} A.~V.,  {Ergma} E.~V.,  1979, Soviet Astronomy Letters, \href
  {http://adsabs.harvard.edu/abs/1979SvAL....5..284T} {5, 284}

\bibitem[\protect\citeauthoryear{{Tutukov} \& {Yungel'Son}}{{Tutukov} \&
  {Yungel'Son}}{1972}]{ty72}
{Tutukov} A.~V.,  {Yungel'Son} L.~R.,  1972, \mn@doi [Astrophysics]
  {10.1007/BF01011356}, \href
  {http://adsabs.harvard.edu/abs/1972Ap......8..227T} {8, 227}

\bibitem[\protect\citeauthoryear{{van Haaften}, {Nelemans}, {Voss}, {Toonen},
  {Portegies Zwart}, {Yungelson}  \& {van der Sluys}}{{van Haaften}
  et~al.}{2013}]{vnvt+13}
{van Haaften} L.~M.,  {Nelemans} G.,  {Voss} R.,  {Toonen} S.,  {Portegies
  Zwart} S.~F.,  {Yungelson} L.~R.,   {van der Sluys} M.~V.,  2013, \mn@doi
  [\aap] {10.1051/0004-6361/201220552}, \href
  {http://adsabs.harvard.edu/abs/2013A%26A...552A..69V} {552, A69}

\bibitem[\protect\citeauthoryear{{Webbink}}{{Webbink}}{1984}]{webb84}
{Webbink} R.~F.,  1984, \mn@doi [\apj] {10.1086/161701}, \href
  {http://adsabs.harvard.edu/abs/1984ApJ...277..355W} {277, 355}

\bibitem[\protect\citeauthoryear{{Webbink}}{{Webbink}}{1988}]{webb88}
{Webbink} R.~F.,  1988, {The Formation and Evolution of Symbiotic Stars}.
p.~311

\bibitem[\protect\citeauthoryear{{Williams} \& {Shafter}}{{Williams} \&
  {Shafter}}{2004}]{ws04}
{Williams} S.~J.,  {Shafter} A.~W.,  2004, \mn@doi [\apj] {10.1086/422833},
  \href {http://adsabs.harvard.edu/abs/2004ApJ...612..867W} {612, 867}

\bibitem[\protect\citeauthoryear{{Williams}, {Darnley}, {Bode}, {Keen}  \&
  {Shafter}}{{Williams} et~al.}{2014}]{wdbk+14}
{Williams} S.~C.,  {Darnley} M.~J.,  {Bode} M.~F.,  {Keen} A.,   {Shafter}
  A.~W.,  2014, \mn@doi [\apjs] {10.1088/0067-0049/213/1/10}, \href
  {http://adsabs.harvard.edu/abs/2014ApJS..213...10W} {213, 10}

\bibitem[\protect\citeauthoryear{{Williams}, {Darnley}, {Bode}  \&
  {Shafter}}{{Williams} et~al.}{2016}]{wdbs16}
{Williams} S.~C.,  {Darnley} M.~J.,  {Bode} M.~F.,   {Shafter} A.~W.,  2016,
  \mn@doi [\apj] {10.3847/0004-637X/817/2/143}, \href
  {http://adsabs.harvard.edu/abs/2016ApJ...817..143W} {817, 143}

\bibitem[\protect\citeauthoryear{{Wolf}, {Bildsten}, {Brooks}  \&
  {Paxton}}{{Wolf} et~al.}{2013}]{wbbp13}
{Wolf} W.~M.,  {Bildsten} L.,  {Brooks} J.,   {Paxton} B.,  2013, \mn@doi
  [\apj] {10.1088/0004-637X/777/2/136}, \href
  {http://adsabs.harvard.edu/abs/2013ApJ...777..136W} {777, 136}

\bibitem[\protect\citeauthoryear{{Woods} \& {Ivanova}}{{Woods} \&
  {Ivanova}}{2011}]{wi11}
{Woods} T.~E.,  {Ivanova} N.,  2011, \mn@doi [\apjl]
  {10.1088/2041-8205/739/2/L48}, \href
  {http://adsabs.harvard.edu/abs/2011ApJ...739L..48W} {739, L48}

\bibitem[\protect\citeauthoryear{{Yaron}, {Prialnik}, {Shara}  \&
  {Kovetz}}{{Yaron} et~al.}{2005}]{ypsk05}
{Yaron} O.,  {Prialnik} D.,  {Shara} M.~M.,   {Kovetz} A.,  2005, \mn@doi
  [\apj] {10.1086/428435}, \href
  {http://adsabs.harvard.edu/abs/2005ApJ...623..398Y} {623, 398}

\bibitem[\protect\citeauthoryear{{Yungelson}, {Livio}, {Truran}, {Tutukov}  \&
  {Fedorova}}{{Yungelson} et~al.}{1996}]{yltt+96}
{Yungelson} L.,  {Livio} M.,  {Truran} J.~W.,  {Tutukov} A.,   {Fedorova} A.,
  1996, \mn@doi [\apj] {10.1086/177562}, \href
  {http://adsabs.harvard.edu/abs/1996ApJ...466..890Y} {466, 890}

\bibitem[\protect\citeauthoryear{{Yungelson}, {Livio}  \&
  {Tutukov}}{{Yungelson} et~al.}{1997}]{ylt97}
{Yungelson} L.,  {Livio} M.,   {Tutukov} A.,  1997, \apj, \href
  {http://adsabs.harvard.edu/abs/1997ApJ...481..127Y} {481, 127}

\bibitem[\protect\citeauthoryear{{Zorotovic}, {Schreiber}, {G{\"a}nsicke}  \&
  {Nebot G{\'o}mez-Mor{\'a}n}}{{Zorotovic} et~al.}{2010}]{zsgn10}
{Zorotovic} M.,  {Schreiber} M.~R.,  {G{\"a}nsicke} B.~T.,   {Nebot
  G{\'o}mez-Mor{\'a}n} A.,  2010, \mn@doi [\aap] {10.1051/0004-6361/200913658},
  \href {http://adsabs.harvard.edu/abs/2010A%26A...520A..86Z} {520, A86}




\makeatother
\end{thebibliography}


\label{lastpage}
\end{document}